\documentclass{rspublic}

\begin{document}

\title[AKLT Density Matrix]{Entanglement and Density Matrix \\
of a Block of Spins in AKLT Model}

\author[Y. Xu, H. Katsura, T. Hirano \& V. Korepin]{Ying Xu${}^{1}$, Hosho Katsura${}^{2}$, Takaaki Hirano${}^{2}$ \\ \& Vladimir E. Korepin${}^{1}$}

\affiliation{${}^{1}$C.N. Yang Institute for Theoretical Physics \\
 State University of New York at Stony Brook, Stony Brook, NY 11794-3840, USA
 \\
 ${}^{2}$Department of Applied Physics \\ 
 The University of Tokyo, 7-3-1 Hongo, Bunkyo-ku, Tokyo 113-8656, Japan}

%%%%%%%%%%%%%%%%%%%%%%%%%%%%%%%%%%%%%%%%%%%%%%%%%%%%%%%%%%%%%%%%%%%%%%%%%%%%%%%%%%%%%%%%%%%%%%%%%%%%%%%%%%%%%%%%%%%%%%%%%%%%%%%%%%%%%%%%%%%%%%%%%%%%%%%%%%%%%%%%%%%%%%%%%%%%%%%%%%%%%%%%%%%%%%%%
\label{firstpage}

\maketitle

\begin{abstract}{AKLT, Density matrix, Entanglement, Valence Bond Solid}
We study a 1-dimensional AKLT spin chain, consisting of spins $S$ in the bulk and $S/2$ at both ends. The unique ground state of this AKLT model is described by the Valence-Bond-Solid (VBS) state. We investigate the density matrix of a contiguous block of bulk spins in this ground state. It is shown that the density matrix is a projector onto a subspace of dimension $\left(S+1\right)^{2}$. This subspace is described by non-zero eigenvalues and corresponding eigenvectors of the density matrix. We prove that for large block the von Neumann entropy coincides with Renyi entropy and is equal to $\ln\left(S+1\right)^{2}$.
\end{abstract}

%%%%%%%%%%%%%%%%%%%%%%%%%%%%%%%%%%%%%%%%%%%%%%%%%%%%%%%%%%%%%%%%%%%%%%%%%%%%%%%%%%%%%%%%%%%%%%%%%%%%%%%%%%%%%%%%%%%%%%%%%%%%%%%%%%%%%%%%%%%%%%%%%%%%%%%%%%%%%%%%%%%%%%%%%%%%%%%%%%%%%%%%%%%%%%%%
\section{Introduction}
\label{intro}

There is considerable current interest in studying various interacting quantum systems from the quantum information perspective. Quantum entanglement is a fundamental measure of how much quantum effects we can observe and use to control one quantum system by another, and it is the primary resource in quantum computation and quantum information processing (Bennett \& DiVincenzo 2000, Lloyd 1993). Entanglement properties also play an important role in condensed matter physics, such as phase transitions (Osterloh, \textit{et al.} 2002; Osborne \& Nielsen 2002) and macroscopic properties of solids (Ghosh, \textit{et al.} 2003; Vedral 2004). 
Extensive research has been undertaken to understand quantum entanglement for spin chains, correlated electrons, interacting bosons as well as other models, see Amico, \textit{et al.} (2007), Audenaert, \textit{et al.} (2002), Fan \& Korepin (2008), Katsura, \textit{et al.} (2007$b$), Fan, \textit{et al.} (2007), %Osterloh, \textit{et al.}, Osborne \& Nielsen (2002), Ghosh, \textit{et al.} (2003), Vedral (2004), 
Arnesen, \textit{et al.} (2001), Korepin (2004), Verstraete, \textit{et al.} (2004$a$, $b$), Campos Venuti, \textit{et al.} (2006), Jin \& Korepin (2004), Vedral (2004), Latorre, \textit{et al.} (2004$a$, $b$, 2005), Orus (2005), 
Orus \& Latorre (2004), Pachos \& Plenio (2004), Plenio \textit{et al.} (2004), Fan \& Lloyd (2005), Chen, \textit{et al.} (2004), Zanardi \& Rasetti (1999), Popkov \& Salerno (2004), Keating \& Mezzadri (2004), Gu, \textit{et al.} (2003, 2004), Wang, \textit{et al.} (2004), Wang \& Kais (2004), Holzhey, \textit{et al.} (1994), Calabrese \& Cardy (2004), Levin \& Wen (2006), Kitaev \& Preskill (2006), Ryu \& Hatsugai (2006), %Hatsugai (2005), 
Hirano \& Hatsugai (2007) for reviews and references.  Characteristic functions of quantum entanglement, such as von Neumann entropy and Renyi entropy, are obtained and discussed through studying reduced density matrices of subsystems (Fan, \textit{et al.} 2004; Katsura, \textit{et al.} 2007$a$; Its, \textit{et al.} 2005; Franchini, \textit{et al.} 2007, 2008; Vidal, \textit{et al.} 2003). An area law for the von Neumann entropy in harmonic lattice systems has been extensively studied (Cramer, \textit{et al.} 2006, 2007; Plenio, \textit{et al.} 2005).

In this paper we study a spin chain model introduced by Affleck, Kennedy, Lieb and Tasaki (AKLT) (Affleck, \textit{et al.} 1987, 1988). We consider a $1$-dimensional AKLT model with spin-$S$ in the bulk and spin-$S/2$ at both ends. The ground state of this model is a unique pure state (Arovas, \textit{et al.} 1988). It is known as the Valence-Bond-Solid (VBS) state, which plays a significant role in condensed matter physics. The VBS state is closely related to Laughlin ansatz (Laughlin 1983, Iblisdir \textit{et al.} 2007) and fractional quantum Hall effect (Arovas, \textit{et al.} 1988). It enables us to understand ground state properties of anti-ferromagnetic integer-spin chains where the finite energy gap known as the Haldane gap exists (Haldane 1983). Universal quantum computation based on VBS states (Verstraete \& Cirac 2004) and an implementation of the AKLT hamiltonian in optical lattices (Garcia-Ripoll \textit{et al.} 2004) have also been proposed.

The density matrix of a contiguous block of bulk spins (we call it \textit{the density matrix} later for short) has been studied extensively in Kirillov \& Korepin (1990), Verstraete, \textit{et al.} (2004$a$), Fan, \textit{et al.} (2004), Katsura, \textit{et al.} (2007$a$), Freitag \& Muller-Hartmann (1991). It contains information of all correlation functions (Jin \& Korepin 2004, Katsura, \textit{et al.} 2007$a$, Arovas, \textit{et al.} 1988). Moreover, it has been shown in Fan, \textit{et al.} (2004), Katsura, \textit{et al.} (2007$a$) that the density matrix is independent of the size of the chain and the location of the block relative to the ends. Therefore we can take the length of the block equal to the length of the whole chain. (i.e. we can add two ending spins $S/2$ directly to the block.) Then by using the Schmidt decomposition (Nielsen \& Chuang 2000), we can show that the density matrix of the block is equivalent to the density matrix of the two ending spins. By equivalent we mean that all non-zero eigenvalues are the same. Using this method, eigenvalues of the density matrix as well as entanglement entropies were obtained (Fan, \textit{et al.} 2004; Katsura, \textit{et al.} 2007$a$; Freitag \& Muller-Hartmann 1991) without knowing the eigenvectors explicitly. 

However, eigenvectors of the density matrix have their own importance. They can be used to study the structure and symmetries of the density matrix explicitly both for finite block and in large block limit. The construction of eigenvectors also provides us with a possible method to diagonalize the density matrix directly. As to be shown in following sections (see \S\ref{sec1}$\,\ref{sec1.3}\,$, \S\ref{sec1}$\,\ref{sec1.4}\,$, \S\ref{sec3}$\,\ref{sec3.3}\,$), the eigenvectors also have their own physical meaning as degenerate zero-energy ground states. In the context of the Haldane gap, these degenerate states are known as \textit{edge states} and have been observed in the $S=1$ spin chain compound (Hagiwara \textit{et al.} 1990). Furthermore, eigenvectors become indispensable in quantum computing algorithms, particularly in discussing quantum measurements. 

In this paper, we consider AKLT models with two different boundary conditions. Let's first take spin $S=1$ for example. The system consists of a linear chain of $N$ spin-$1$'s in the bulk, and two spin-$1/2$'s on the boundaries. We shall denote by $\boldsymbol{S}_{j}$ the vector spin operator at site $j$ ($j=0,1,\ldots,N+1$). The Hamiltonian is
\begin{eqnarray}
	H_{uniq}=\frac{1}{2}\sum^{N-1}_{j=1} \left(\boldsymbol{S}_{j}\cdot\boldsymbol{S}_{j+1}+\frac{1}{3}\left(\boldsymbol{S}_{j}\cdot\boldsymbol{S}_{j+1}\right)^{2}+\frac{2}{3}\right)+\pi_{0,1}+\pi_{N, N+1}. \label{uniq1}
\end{eqnarray}
The boundary terms $\pi$ describe interactions of a spin-$1/2$ and a spin-$1$. Each term is a projector onto a state with spin $3/2$:
\begin{eqnarray}
	\pi_{0,1}\equiv\frac{2}{3}\left(1+\boldsymbol{S}_{0}\cdot\boldsymbol{S}_{1}\right),  \qquad   \pi_{N,N+1}\equiv\frac{2}{3}\left(1+\boldsymbol{S}_{N}\cdot\boldsymbol{S}_{N+1}\right). \label{boun1}
\end{eqnarray}
The Hamiltonian (\ref{uniq1}) has a unique ground state (VBS state), thus we shall call it \textit{the unique Hamiltonian}. Alternatively, if we consider spin-$1$'s at every site including the boundaries, then the Hamiltonian takes the form
\begin{eqnarray}
	H_{deg}=\frac{1}{2}\sum^{N-1}_{j=1} \left(\boldsymbol{S}_{j}\cdot\boldsymbol{S}_{j+1}+\frac{1}{3}\left(\boldsymbol{S}_{j}\cdot\boldsymbol{S}_{j+1}\right)^{2}+\frac{2}{3}\right). \label{dege1}
\end{eqnarray}
The ground states of this Hamiltonian are $4$-fold degenerate. We shall call (\ref{dege1}) \textit{the degenerate Hamiltonian}.

For generic spin-$S$, the unique Hamiltonian is
\begin{eqnarray}
	H_{uniq}=\sum^{N-1}_{j=1}\sum^{2S}_{J=S+1} C_{J}P^{J}_{j, j+1}+\pi_{0,1}+\pi_{N, N+1}, \label{uniq}
\end{eqnarray}
where the projector $P^{J}_{j, j+1}$ projects the bond spin $\boldsymbol{J}_{j, j+1}\equiv\boldsymbol{S}_{j}+\boldsymbol{S}_{j+1}$ onto the subspace with total spin $J$ ($J=S+1, \ldots, 2S$). The boundary terms describe interactions between a spin-$S/2$ and a spin-$S$:
\begin{eqnarray}
	\pi_{0,1}\equiv\sum^{3S/2}_{J=S/2+1} D_{J}P^{J}_{0, 1}, \qquad    \pi_{N,N+1}\equiv\sum^{3S/2}_{J=S/2+1} D_{J}P^{J}_{N, N+1}. \label{boun}
\end{eqnarray}
Both coefficients $C_{J}$ and $D_{J}$ can take arbitrary positive values. Correspondingly, the degenerate Hamiltonian with spin-$S$ at every site takes the form
\begin{eqnarray}
	H_{deg}=\sum^{N-1}_{j=1}\sum^{2S}_{J=S+1} C_{J}P^{J}_{j, j+1}. \label{dege}
\end{eqnarray}
The degeneracy of the ground states is $\left(S+1\right)^{2}$. This will be important in description of eigenvectors of the density matrix (see \S\ref{sec1}$\,\ref{sec1.4}\,$ and \S\ref{sec2}).

Consider the AKLT spin chain system with the unique Hamiltonian (\ref{uniq}) in the VBS ground state. The density matrix $\boldsymbol{\rho}$ of the whole chain is a projector onto the unique VBS ground state (see (\ref{pure})). If we pick up a block of $L$ contiguous bulk spins as a subsystem and trace out all degrees of freedom outside the block, then we obtain the density matrix $\boldsymbol{\rho}_{L}$ of the subsystem (see (\ref{trac})). Because of entanglement with spins outside the block, $\boldsymbol{\rho}_{L}$ will no longer be a pure state density matrix as $\boldsymbol{\rho}$ is in general. We shall prove that the density matrix $\boldsymbol{\rho}_{L}$ is a projector onto a $(S+1)^{2}$-dimensional subspace of the complete Hilbert space associated with the block (see \S\ref{sec1}$\,\ref{sec1.4}\,$ and \S\ref{sec2}). The degenerate Hamiltonian (\ref{dege}) becomes essential in description of this subspace. When the degenerate Hamiltonian has its size $N$ equal to that of the block $L$, it is referred to as \textit{the block Hamiltonian} and denoted by $H_{b}$ which is defined by (\ref{degel}). It turns out that the block Hamiltonian $H_{b}$ (i.e. the degenerate Hamiltonian $H_{deg}$ in (\ref{dege}) with $N=L$) defines the density matrix $\boldsymbol{\rho}_{L}$ completely in the large block limit $L\rightarrow\infty$. The zero-energy ground states of the block Hamiltonian $H_{b}$ span the subspace that the density matrix $\boldsymbol{\rho}_{L}$ projects onto. So that $\boldsymbol{\rho}_{L}$ can be represented as the zero-temperature limit of the canonical ensemble density matrix defined by $H_{b}$:
\begin{eqnarray}
	\boldsymbol{\rho}_{L}=\lim_{\beta\rightarrow +\infty}\frac{\re^{-\beta H_{b}}}{Tr\left[\re^{-\beta H_{b}}\right]}, \qquad L\rightarrow\infty, \label{enlim}
\end{eqnarray}
where 
\begin{eqnarray}
H_b \equiv H_{deg}(\mbox{with}\ N=L)=\sum^{L-1}_{j=1}\sum^{2S}_{J=S+1} C_{J}P^{J}_{j, j+1}. \label{degel_former}
\end{eqnarray}
In the zero-temperature limit, contributions from excited states of $H_{b}$ all vanish and the right hand side of (\ref{enlim}) turns into a projector onto the ground states of the block Hamiltonian.

As main subjects of the paper, we will construct eigenvectors and derive expressions for corresponding eigenvalues of the density matrix. We will show that the density matrix is a projector.
%We shall prove that the density matrix is a projector onto a $(S+1)^{2}$-dimensional subspace of the complete Hilbert space of the block.
The paper is divided into four parts:
\begin{enumerate}
	\item We calculate the density matrix, prove a theorem on eigenvectors and express eigenvalues in two different forms using the Schwinger representation (\S\ref{sec1}).
	\item We investigate the structure of the density matrix in the large block limit. As characteristic functions of quantum entanglement, the von Neumann entropy and the Renyi entropy are obtained in the limit (\S\ref{sec2}).  
	\item We study the density matrix using a different representation (a pure algebraic method) for spin $S=1$ (\S\ref{sec3}).
	\item An alternative proof of the theorem on eigenvectors is given as we take a different approach (\S\ref{sec4}).
\end{enumerate}

%%%%%%%%%%%%%%%%%%%%%%%%%%%%%%%%%%%%%%%%%%%%%%%%%%%%%%%%%%%%%%%%%%%%%%%%%%%%%%%%%%%%%%%%%%%%%%%%%%%%%%%%%%%%%%%%%%%%%%%%%%%%%%%%%%%%%%%%%%%%%%%%%%%%%%%%%%%%%%%%%%%%%%%%%%%%%%%%%%%%%%%%%%%%%%%%
\section{Density Matrix for Generic Spin-$S$}
\label{sec1}

%%%%%%%%%%%%%%%%%%%%%%%%%%%%%%%%%%%%%%%%%%%%%%%%%%%%%%%%%%%%%%%%%%%%%%%%%%%%%%%%%%%%%%%%%%%%%%%%
\subsection{Ground State of the Unique Hamiltonian}
\label{sec1.1}

We start with the ground state of the unique Hamiltonian (\ref{uniq}). It is given in the Schwinger representation by the VBS state (Arovas, \textit{et al.} 1988)
\begin{eqnarray}
	|\mbox{VBS}\rangle \equiv
 \prod^{N}_{j=0}
\left(a^{\dagger}_{j}b^{\dagger}_{j+1}-b^{\dagger}_{j}a^{\dagger}_{j+1}\right)^{S}|\mbox{vac}\rangle, \label{vbs}
\end{eqnarray}
where $a^{\dagger}$, $b^{\dagger}$ are bosonic creation operators and $\left|\mbox{vac}\right\rangle$ is destroyed by any of the annihilation operators $a$, $b$. These operators satisfy $[a_{i}, a^{\dagger}_{j}]=[b_{i}, b^{\dagger}_{j}]=\delta_{ij}$ with all other commutators vanishing. The spin operators are represented as $S^{+}_{j}=a^{\dagger}_{j}b_{j}$, $S^{-}_{j}=b^{\dagger}_{j}a_{j}$, $S^{z}_{j}=(a^{\dagger}_{j}a_{j}-b^{\dagger}_{j}b_{j})/2$. To reproduce the dimension of the spin-$S$ Hilbert space at each site, an additional constraint on the total boson occupation number is required, namely $(a^{\dagger}_{j}a_{j}+b^{\dagger}_{j}b_{j})/2=S$. More details and properties of the VBS state in the Schwinger representation can be found in Kirillov \& Korepin (1990), Arovas, \textit{et al.} (1988), Auerbach (1998). The pure state density matrix of the VBS ground state (\ref{vbs}) is
\begin{eqnarray}
	\boldsymbol{\rho}=\frac{|\mbox{VBS}\rangle\langle \mbox{VBS}|}{\langle \mbox{VBS}|\mbox{VBS}\rangle}. \label{pure}
\end{eqnarray}
For normalization $\langle \mbox{VBS}|\mbox{VBS}\rangle$ of the VBS state, see \ref{secA1}.

%%%%%%%%%%%%%%%%%%%%%%%%%%%%%%%%%%%%%%%%%%%%%%%%%%%%%%%%%%%%%%%%%%%%%%%%%%%%%%%%%%%%%%%%%%%%%%%%
\subsection{Density Matrix of a Block of Bulk Spins}
\label{sec1.2}

We take a block of $L$ contiguous bulk spins as a subsystem. In order to calculate the density matrix of the block, it is convenient to introduce a spin coherent state representation. We introduce spinor coordinates
\begin{eqnarray}
	\left(u, v\right)\equiv\left(\cos\frac{\theta}{2}\re^{\ri\frac{\phi}{2}}, \sin\frac{\theta}{2}\re^{-\ri\frac{\phi}{2}}\right), \qquad 0\leq\theta\leq\pi, \quad 0\leq\phi\leq 2\pi. \label{spin}
\end{eqnarray}
Then for a point $\hat{\Omega}\equiv (\sin\theta\cos\phi, \sin\theta\sin\phi, \cos\theta)$ on the unit sphere, the spin-$S$ coherent state is defined as
\begin{eqnarray}
	|\hat{\Omega}\rangle\equiv\frac{\left(ua^{\dagger}+vb^{\dagger}\right)^{2S}}{\sqrt{\left(2S\right)!}}|\mbox{vac}\rangle. \label{cohe}
\end{eqnarray}
Here we have fixed the overall phase (a $U(1)$ gauge degree of freedom) since it has no physical content. Note that (\ref{cohe}) is covariant under $SU(2)$ transforms (see \S\ref{sec2}). The set of coherent states is complete (but not orthogonal) such that (Freitag \& Muller-Hartmann 1991; Arecchi, \textit{et al.} 1972)
\begin{eqnarray}
	\frac{2S+1}{4\pi}\int \rd\hat{\Omega} |\hat{\Omega}\rangle \langle \hat{\Omega}|=\sum^{S}_{m=-S}|S, m\rangle\langle S, m|=I_{2S+1}, \label{comp}
\end{eqnarray}
where $|S, m\rangle$ denote the eigenstate of $\boldsymbol{S}^{2}$ and $S_{z}$, and $I_{2S+1}$ is the identity of the $(2S+1)$-dimensional Hilbert space for spin-$S$. The completeness relation (\ref{comp}) can be used in taking trace of an arbitrary operator.

Now we calculate the density matrix of a block of $L$ contiguous bulk spins in the VBS state (\ref{vbs}). By definition, this is achieved by taking the pure state density matrix  (\ref{pure}) and tracing out all spin degrees of freedom outside the block:
\begin{eqnarray}
	\boldsymbol{\rho}_{L}\equiv Tr_{0, 1, \ldots, k-1, k+L, \ldots, N, N+1}\ \left[\boldsymbol{\rho}\right],\qquad 1\leq k,\quad k+L-1\leq N. \label{trac}
\end{eqnarray}
Here the block of length $L$ starts from site $k$ and ends at site $k+L-1$. $\boldsymbol{\rho}_{L}$ is no longer a pure state density matrix because of entanglement of the block with the environment (sites outside the block of the spin chain). It was shown in Section 2 of Jin \& Korepin (2004) that entries of the density matrix are multi-point correlation functions in the ground state. The original proof was for spin $S=1/2$. This statement is generalized to generic spin-$S$ in \ref{secA4}. 

Using the coherent state representation (\ref{cohe}) and completeness relation (\ref{comp}), $\boldsymbol{\rho}_{L}$ can be written as (Katsura, \textit{et al.} 2007$a$)
\begin{eqnarray}
	&&\boldsymbol{\rho}_{L}=
	\label{roug} \\
	&&\frac{
\displaystyle\int\left[\prod^{k-1}_{j=0}\prod^{N+1}_{j=k+L}\rd\hat{\Omega}_{j}\right]\prod^{k-2}_{j=0}\prod^{N}_{j=k+L}\left[\frac{1}{2}(1-\hat{\Omega}_{j}\cdot\hat{\Omega}_{j+1})\right]^{S}
	B^{\dagger}|\mbox{VBS}_{L}\rangle\langle \mbox{VBS}_{L}|B}
	{\displaystyle\left[\frac{(2S+1)!}{4\pi}\right]^{L}
\int\left[\prod^{N+1}_{j=0}\rd\hat{\Omega}_{j}\right]\prod^{N}_{j=0}\left[\frac{1}{2}(1-\hat{\Omega}_{j}\cdot\hat{\Omega}_{j+1})\right]^{S}}. \nonumber 
\end{eqnarray}
Here the boundary operator $B$ and block VBS state $\left|\mbox{VBS}_{L}\right\rangle$ are defined as
\begin{eqnarray}
	&&B\equiv \left(u_{k-1}b_{k}-v_{k-1}a_{k}\right)^{S}\left(a_{k+L-1}v_{k+L}-b_{k+L-1}u_{k+L}\right)^{S},\label{bope} \\
	&&|\mbox{VBS}_{L}\rangle \equiv
 \prod^{k+L-2}_{j=k}
\left(a^{\dagger}_{j}b^{\dagger}_{j+1}-b^{\dagger}_{j}a^{\dagger}_{j+1}\right)^{S}|\mbox{vac}\rangle, \label{vbsl}
\end{eqnarray}
respectively. Note that both $B$ and $|\mbox{VBS}_{L}\rangle$ are $SU(2)$ covariant (see \S\ref{sec2}). The expression (\ref{roug}) can be simplified. We can perform the integrals over $\hat{\Omega}_{j}$ ($j=0, 1, \ldots, k-2, k+L+1, \ldots, N, N+1$) in the numerator and all integrals in the denominator (see \ref{secA1}). After integrating over these variables, the density matrix $\boldsymbol{\rho}_{L}$ turns out to be independent of both the starting site $k$ and the total length $L$ of the block. This property has been proved in Fan, \textit{et al.} (2004) for spin $S=1$ (using a different representation, namely the maximally entangled states, see \S\ref{sec3}) and generalized in Katsura, \textit{et al.} (2007$a$) for generic spin-$S$. Therefore, we can choose $k=1$ and the density matrix takes the form
\begin{eqnarray}
	\boldsymbol{\rho}_{L}=\left[\frac{S+1}{(2S+1)!}\right]^{L}\frac{(S+1)}{(4\pi)^{2}}
	\int \rd\hat{\Omega}_{0}\rd\hat{\Omega}_{L+1}B^{\dagger}|\mbox{VBS}_{L}\rangle\langle \mbox{VBS}_{L}|B \label{matr}
\end{eqnarray}
with
\begin{eqnarray}
&&B^{\dagger}=\left(u^{\ast}_{0}b^{\dagger}_{1}-v^{\ast}_{0}a^{\dagger}_{1}\right)^{S}\left(a^{\dagger}_{L}v^{\ast}_{L+1}-b^{\dagger}_{L}u^{\ast}_{L+1}\right)^{S}, \label{bope1} \\
	&&|\mbox{VBS}_{L}\rangle=\prod^{L-1}_{j=1}
\left(a^{\dagger}_{j}b^{\dagger}_{j+1}-b^{\dagger}_{j}a^{\dagger}_{j+1}\right)^{S}|\mbox{vac}\rangle. \label{vbsl1}	
\end{eqnarray}
The last two integral of (\ref{matr}) can be performed, but we keep its present form for later use.

%%%%%%%%%%%%%%%%%%%%%%%%%%%%%%%%%%%%%%%%%%%%%%%%%%%%%%%%%%%%%%%%%%%%%%%%%%%%%%%%%%%%%%%%%%%%%%%%
\subsection{Ground States of the Block Hamiltonian}
\label{sec1.3}

In order to describe the eigenvectors of the density matrix (\ref{matr}), we first study the zero-energy ground states of the degenerate Hamiltonian defined in (\ref{dege}). We choose the length of the spin chain equal to that of the block, i.e. $N=L$, then the degenerate Hamiltonian is called the block Hamiltonian and reads
\begin{eqnarray}
	H_{b}\equiv H_{deg}(\mbox{with}\ N=L)=\sum^{L-1}_{j=1}\sum^{2S}_{J=S+1} C_{J}P^{J}_{j, j+1}. \label{degel}
\end{eqnarray}
Now we define a set of $S+1$ operators covariant under $SU(2)$
\begin{eqnarray}
 A^{\dagger}_{J}\equiv
\left(ua^{\dagger}_{1}+vb^{\dagger}_{1}\right)^{J}  \left(a^{\dagger}_{1}b^{\dagger}_{L}-b^{\dagger}_{1}a^{\dagger}_{L}\right)^{S-J}
\left(ua^{\dagger}_{L}+vb^{\dagger}_{L}\right)^{J}, \quad 0\leq J\leq S. \label{aope}
\end{eqnarray}
These operators act on the direct product of Hilbert spaces of spins at site $1$ and site $L$. Then the set of ground states of (\ref{degel}) can be chosen as
\begin{eqnarray}
	|\mbox{G}; J, \hat{\Omega}\rangle \equiv
A^{\dagger}_{J}|\mbox{VBS}_{L}\rangle, \qquad J=0, \ldots, S. \label{eige}
\end{eqnarray}
Any state $|\mbox{G}; J, \hat{\Omega}\rangle$ of this set for fixed $J$ and $\hat{\Omega}$ is a zero-energy ground state of (\ref{degel}). To prove this we need only to verify: (i) the total power of $a^{\dagger}_{1}$ and $b^{\dagger}_{1}$ is $2S$, so that we have spin-$S$ at the first site; (ii) $-S\leq J^{z}_{1,2}\equiv S^{z}_{1}+S^{z}_{2}\leq S$ by a binomial expansion, so that the maximum value of the bond spin $J_{1,2}$ is $S$ (from $SU(2)$ invariance, see Arovas, \textit{et al.} 1988). These properties are true for any other site $j$ and bond $(j, j+1)$, respectively. Therefore, the state $|\mbox{G}; J, \hat{\Omega}\rangle$ defined in (\ref{eige}) has spin-$S$ at each site and no projection onto the $J_{j, j+1}>S$ subspace for any bond.

The set of states $\{|\mbox{G}; J, \hat{\Omega}\rangle\}$ depend on a discrete parameter $J$ as well as a continuous unit vector $\hat{\Omega}$. States with the same $J$ value are not orthogonal. The rank of a set of states with the same $J$ value is $2J+1$, which can be obtained from the completeness relation (\ref{compga}) (see \ref{secA2} and Hamermesh 1989). Thus the total number of linearly independent states of the set $\{|\mbox{G}; J, \hat{\Omega}\rangle\}$ is $\sum^{S}_{J=0}(2J+1)=(S+1)^2$, which is exactly the degeneracy of the ground states of (\ref{degel}). So that $\{|\mbox{G}; J, \hat{\Omega}\rangle\}$ forms a complete set of zero-energy ground states.  

We also introduce an orthogonal basis in description of the degenerate zero-energy ground states. It is shown in \ref{secA2} and Hamermesh (1989) that $A^{\dagger}_{J}$ (\ref{aope}) can be expanded in terms of spin creation operators $\Psi^{\dagger}_{JM}$ ($M=-J,\ldots,J$) defined in (\ref{crea2}). Operator $\Psi^{\dagger}_{JM}$ acts on the direct product of two Hilbert spaces of spins at site $1$ and site $L$ (\ref{crea3}) and can be expressed in terms of bosonic creation operators in the Schwinger representation (\ref{2def}). If we define a set of \textit{degenerate VBS states} $\{|\mbox{VBS}_L(J,M)\rangle\}$ such that
\begin{eqnarray}
	|\mbox{VBS}_L(J,M)\rangle\equiv \Psi^{\dagger}_{JM}|\mbox{VBS}_L\rangle, \quad J=0,...,S, \quad M=-J, ...,J, \label{devb}
\end{eqnarray}
then these $(S+1)^{2}$ states (\ref{devb}) are not only linearly independent but also mutually orthogonal (\ref{secA3}). Furthermore, any ground state $|\mbox{G}; J, \hat{\Omega}\rangle$ can be written as a linear superposition over these degenerate VBS states, and \textit{vice versa} (see (\ref{line}) of \ref{secA2}). The set $\{|\mbox{VBS}_L(J,M)\rangle\}$ differs from $\{|\mbox{G}; J, \hat{\Omega}\rangle\}$ by a change of basis, so that it also forms a complete set of zero-energy ground states.

%%%%%%%%%%%%%%%%%%%%%%%%%%%%%%%%%%%%%%%%%%%%%%%%%%%%%%%%%%%%%%%%%%%%%%%%%%%%%%%%%%%%%%%%%%%%%%%%
\subsection{Eigenvectors of the Density Matrix}
\label{sec1.4}

Eigenvalues of the density matrix (\ref{matr}) are derived for spin-$1$ in Fan, \textit{et al.} (2004) and for spin-$S$ in Katsura, \textit{et al.} (2007$a$). Because the density matrix is independent of both the total length of 
the spin chain and the starting site of the block, we can add boundary spins directly to the ends of the block. It was shown in Fan, \textit{et al.} (2004), Katsura, \textit{et al.} (2007$a$) by a Schmidt decomposition (Nielsen \& Chuang 2000) that non-zero eigenvalues of the density matrix (\ref{matr}) are equal to those of the density matrix of the two boundary spins. All other eigenvalues of the density matrix (\ref{matr}) are zero. This fact reveals the structure of the density matrix as a projector onto a subspace of dimension $(S+1)^{2}$.

Now we propose a theorem on the eigenvectors of the density matrix given by (\ref{matr}). The explicit construction of eigenvectors allows us to diagonalize the density matrix directly. The set of eigenvectors also spans the subspace that the density matrix projects onto.

\begin{theorem}
\label{theorem1}
	Eigenvectors of the density matrix $\boldsymbol{\rho}_{L}$ (\ref{matr}) with non-zero eigenvalues are given by the set $\{|\mbox{G}; J, \hat{\Omega}\rangle\}$ (\ref{eige}), or, equivalently, by the set \\ $\{|\mbox{VBS}_L(J,M)\rangle\}$ (\ref{devb}). i.e. they are zero-energy ground states of the block Hamiltonian $H_{b}$ (\ref{degel}).
\end{theorem}

We prove the theorem by showing that the density matrix $\boldsymbol{\rho}_{L}$ (\ref{matr}) can be written as a projector in diagonal form onto the orthogonal degenerate VBS states $\{|\mbox{VBS}_{L}(J, M)\rangle\}$ introduced in (\ref{devb}). An alternative proof taking a different approach is given in \S\ref{sec4}.

First, it is realized from the definition of spinor coordinates (\ref{spin}) that if we change variables $(u, v)$ to $(\ri v^{\ast}, -\ri u^{\ast})$, then the unit vector $\hat{\Omega}$ is inverted about the origin to $-\hat{\Omega}$. So that we have (Katsura, \textit{et al.} 2007$a$)
\begin{eqnarray}
	(u^{\ast}b^{\dagger}-v^{\ast}a^{\dagger})^{S}|\mbox{vac}\rangle=\ri^{S}\sqrt{S!} \ |-\hat{\Omega}\rangle, \label{inve}
\end{eqnarray}
where $|-\hat{\Omega}\rangle$ means a spin-$S/2$ coherent state for a point opposite to $\hat{\Omega}$ on the unit sphere. Therefore, taking expressions of the boundary operator $B^{\dagger}$ (\ref{bope1}) and the block VBS state $|\mbox{VBS}_{L}\rangle$ (\ref{vbsl1}), we have
\begin{eqnarray}
	&&B^{\dagger}|\mbox{VBS}_{L}\rangle \label{bvbs}= \\
	&&S!\prod^{L-1}_{j=1}
\left(a^{\dagger}_{j}b^{\dagger}_{j+1}-b^{\dagger}_{j}a^{\dagger}_{j+1}\right)^{S}|-\hat{\Omega}_{0}\rangle_{1}\otimes|\mbox{vac}\rangle_{2}\otimes\cdots\otimes|\mbox{vac}\rangle_{L-1}\otimes|-\hat{\Omega}_{L+1}\rangle_{L}. \nonumber
\end{eqnarray}
Consequently the density matrix $\boldsymbol{\rho}_{L}$ (\ref{matr}) can be re-written as
\begin{eqnarray}
	\boldsymbol{\rho}_{L}&&=\left[\frac{S+1}{(2S+1)!}\right]^{L}\frac{S!S!}{S+1}\prod^{L-1}_{j=1}
\left(a^{\dagger}_{j}b^{\dagger}_{j+1}-b^{\dagger}_{j}a^{\dagger}_{j+1}\right)^{S} \label{rewr} \\
&&\cdot I^{(1)}_{S+1}\otimes|\mbox{vac}\rangle_{2}\langle \mbox{vac}|\otimes\cdots\otimes|\mbox{vac}\rangle_{L-1}\langle \mbox{vac}|\otimes I^{(L)}_{S+1}\prod^{L-1}_{j=1}\left(a_{j}b_{j+1}-b_{j}a_{j+1}\right)^{S}, \nonumber
\end{eqnarray}
where $I^{(1)}_{S+1}$ and $I^{(L)}_{S+1}$ are $(S+1)$-dimensional identities associated with site $1$ and site $L$, respectively. In obtaining (\ref{rewr}), we have changed integral variables from $\hat{\Omega}_{0}$ , $\hat{\Omega}_{L+1}$ to $-\hat{\Omega}_{0}$, $-\hat{\Omega}_{L+1}$ and performed these two integrals using the completeness relation (\ref{comp}). Next we notice that (see \ref{secA2})
\begin{eqnarray}
	I^{(1)}_{S+1}\otimes I^{(L)}_{S+1}&=&\sum^{S}_{J=0}\sum^{J}_{M=-J}|J, M\rangle_{1,L}\langle J, M| \label{iide} \\
&=&\sum^{S}_{J=0}\sum^{J}_{M=-J}\Psi^{\dagger}_{JM}|\mbox{vac}\rangle_{1}\langle \mbox{vac}|\otimes|\mbox{vac}\rangle_{L}\langle \mbox{vac}|\Psi_{JM}. \nonumber
\end{eqnarray}
As a result, combining (\ref{rewr}) and (\ref{iide}), recalling definitions of $|\mbox{VBS}_{L}\rangle$ (\ref{vbsl1}) and $|\mbox{VBS}_{L}(J, M)\rangle$ (\ref{devb}), the density matrix $\boldsymbol{\rho}_{L}$ takes the following final form
\begin{eqnarray}
	\boldsymbol{\rho}_{L}&=&\left[\frac{S+1}{(2S+1)!}\right]^{L}\frac{S!S!}{S+1}\sum^{S}_{J=0}\sum^{J}_{M=-J}\Psi^{\dagger}_{JM}|\mbox{VBS}_{L}\rangle\langle \mbox{VBS}_{L}|\Psi_{JM} \label{fipr} \\
&\equiv&\left[\frac{S+1}{(2S+1)!}\right]^{L}\frac{S!S!}{S+1}\sum^{S}_{J=0}\sum^{J}_{M=-J}|\mbox{VBS}_{L}(J, M)\rangle\langle \mbox{VBS}_{L}(J, M)|. \nonumber
\end{eqnarray}
The set of degenerate VBS states $\{|\mbox{VBS}_{L}(J, M)\rangle\}$ with $J=0,\ldots,S$ and $M=-J,\ldots,J$ forms an orthogonal basis (see \ref{secA3}). These $(S+1)^{2}$ states also forms a complete set of zero-energy ground states of the block Hamiltonian (\ref{degel}) (see \S\ref{sec1}$\,\ref{sec1.3}\,$ and \S\ref{secA2}). So that in expression (\ref{fipr}) we have put the density matrix as a projector in diagonal form over an orthogonal basis. Each degenerate VBS state $|\mbox{VBS}_{L}(J, M)\rangle$ is an eigenvector of the density matrix, so as any of the state $|\mbox{G}; J, \hat{\Omega}\rangle$ (because of the degeneracy of corresponding eigenvalues of the density matrix, see \S\ref{sec1}$\,\ref{sec1.5}\,$ and \S\ref{sec1}$\,\ref{sec1.6}\,$). Thus we have proved theorem \ref{theorem1}. %\textbf{Theorem 1}.

%%%%%%%%%%%%%%%%%%%%%%%%%%%%%%%%%%%%%%%%%%%%%%%%%%%%%%%%%%%%%%%%%%%%%%%%%%%%%%%%%%%%%%%%%%%%%%%%
\subsection{Eigenvalues of the Density Matrix (Recurrence Formula)}
\label{sec1.5}

Having constructed eigenvectors, we need to specify the corresponding eigenvalues. An explicit expression of eigenvalues is obtained in \S\ref{sec1}$\,\ref{sec1.6}\,$. In this subsection we express eigenvalues through a conjectured recurrence formula as in Katsura, \textit{et al.} (2007$a$) and Freitag \& Muller-Hartmann (1991). Let's apply the density matrix $\boldsymbol{\rho}_{L}$ (\ref{matr}) to the state $|\mbox{G}; J, \hat{\Omega}\rangle$ (\ref{eige}) and get
\begin{eqnarray}
	&&\boldsymbol{\rho}_{L}|\mbox{G}; J, \hat{\Omega}\rangle \label{appl1} \\
	&=&\left[\frac{S+1}{(2S+1)!}\right]^{L}\frac{S+1}{(4\pi)^{2}}
	\int \rd\hat{\Omega}_{0}\rd\hat{\Omega}_{L+1}B^{\dagger}|\mbox{VBS}_{L}\rangle\langle \mbox{VBS}_{L}|BA^{\dagger}_{J}|\mbox{VBS}_{L}\rangle. \nonumber
\end{eqnarray}
Using the coherent state representation (\ref{cohe}) and completeness relation (\ref{comp}), the factor $\langle \mbox{VBS}_{L}|BA^{\dagger}_{J}|\mbox{VBS}_{L}\rangle$
in (\ref{appl1}) can be re-written as
\begin{eqnarray}
	&&\langle \mbox{VBS}_{L}|BA^{\dagger}_{J}|\mbox{VBS}_{L}\rangle \label{appl2} \\
	&=&\displaystyle\left[\frac{(2S+1)!}{4\pi}\right]^{L}\int \left(\prod^{L}_{j=1}\rd\hat{\Omega}_{j}\right)\prod^{L-1}_{j=1}\left[\frac{1}{2}(1-\hat{\Omega}_{j}\cdot\hat{\Omega}_{j+1})\right]^{S}
	\left(u_{0}v_{1}-v_{0}u_{1}\right)^{S} \nonumber \\ &&\cdot\left(uu^{\ast}_{1}+vv^{\ast}_{1}\right)^{J}\left(u^{\ast}_{1}v^{\ast}_{L}-v^{\ast}_{1}u^{\ast}_{L}\right)^{S-J}\left(uu^{\ast}_{L}+vv^{\ast}_{L}\right)^{J}\left(u_{L}v_{L+1}-v_{L}u_{L+1}\right)^{S}. \nonumber
\end{eqnarray}
The factor $\left[\frac{1}{2}(1-\hat{\Omega}_{j}\cdot\hat{\Omega}_{j+1})\right]^{S}$ under the integral of (\ref{appl2}) can be expanded in terms of Legendre polynomials and further in terms of spherical harmonics as (Katsura, \textit{et al.} 2007$a$; Freitag \& Muller-Hartmann 1991)
\begin{eqnarray}
	\left[\frac{1}{2}(1-\hat{\Omega}_{j}\cdot\hat{\Omega}_{j+1})\right]^{S}
	&=&\frac{1}{S+1}\sum^{S}_{l=0}(2l+1)\lambda(l,S)P_{l}(\hat{\Omega}_{j}\cdot\hat{\Omega}_{j+1})\nonumber \\
	&=&\frac{4\pi}{S+1}\sum^{S}_{l=0}\lambda(l,S)\sum^{l}_{m=-l}Y_{lm}(\hat{\Omega}_{j})Y^{\ast}_{lm}(\hat{\Omega}_{j+1}) \label{expa}
\end{eqnarray}
with coefficients $\lambda(l,S)$ given by
\begin{eqnarray}
	\lambda(l,S)\equiv\frac{(-1)^{l}S!(S+1)!}{(S-l)!(S+l+1)!}. \label{lamb}
\end{eqnarray}
Using the expansion (\ref{expa}) and orthogonality of spherical harmonics, the integrals over $\hat{\Omega}_{j}$ with $j=2, \ldots, L-1$ in (\ref{appl2}) can be performed. The result is
\begin{eqnarray}
	\langle \mbox{VBS}_{L}|BA^{\dagger}_{J}|\mbox{VBS}_{L}\rangle &&= \frac{S+1}{(4\pi)^{2}}\left[\frac{(2S+1)!}{S+1}\right]^{L}\sum^{S}_{l=0}(2l+1)\lambda^{L-1}(l,S) \label{resu} \\
	&&\cdot\int \rd\hat{\Omega}_{1}\rd\hat{\Omega}_{L}P_{l}(\hat{\Omega}_{1}\cdot\hat{\Omega}_{L})\left(u_{0}v_{1}-v_{0}u_{1}\right)^{S}\left(uu^{\ast}_{1}+vv^{\ast}_{1}\right)^{J} \nonumber \\
&&\left(u^{\ast}_{1}v^{\ast}_{L}-v^{\ast}_{1}u^{\ast}_{L}\right)^{S-J}\left(uu^{\ast}_{L}+vv^{\ast}_{L}\right)^{J}\left(u_{L}v_{L+1}-v_{L}u_{L+1}\right)^{S}. \nonumber
\end{eqnarray}
We plug the expression (\ref{resu}) into (\ref{appl1}). Using transformation properties under $SU(2)$ and binomial expansion (see \S\ref{sec2}), the integral over $\hat{\Omega}_{0}$ yields that
\begin{eqnarray}
	\int \rd\hat{\Omega}_{0}\left(u^{\ast}_{0}b^{\dagger}_{1}-v^{\ast}_{0}a^{\dagger}_{1}\right)^{S}\left(u_{0}v_{1}-v_{0}u_{1}\right)^{S}=\frac{4\pi}{S+1}\left(u_{1}a^{\dagger}_{1}+v_{1}b^{\dagger}_{1}\right)^{S} \label{int0}
\end{eqnarray}
Similarly we can perform the integral over $\hat{\Omega}_{L+1}$. As a result, the following expression is obtained from (\ref{appl1}):
\begin{eqnarray}
\boldsymbol{\rho}_{L}|\mbox{G}; J, \hat{\Omega}\rangle =  \frac{1}{(4\pi)^{2}}\sum^{S}_{l=0}(2l+1)\lambda^{L-1}(l,S)
K^{\dagger}_{l}(\hat{\Omega})
\left|\mbox{VBS}_{L}\right\rangle. \label{sum}
\end{eqnarray}
The operator $K^{\dagger}_{l}(\hat{\Omega})$ involved in (\ref{sum}) is defined as
\begin{eqnarray}
	K^{\dagger}_{l}(\hat{\Omega})&\equiv&\int \rd\hat{\Omega}_{1} \rd\hat{\Omega}_{L} 
	\left(u_{1}a^{\dagger}_{1}+v_{1}b^{\dagger}_{1}\right)^{S} \left(uu^{\ast}_{1}+vv^{\ast}_{1}\right)^{J}
\left(u^{\ast}_{1}v^{\ast}_{L}-v^{\ast}_{1}u^{\ast}_{L}\right)^{S-J}	
\nonumber \\ &&\cdot
\left(uu^{\ast}_{L}+vv^{\ast}_{L}\right)^{J}
\left(u_{L}a^{\dagger}_{L}+v_{L}b^{\dagger}_{L}\right)^{S}
P_{l}(\hat{\Omega}_{1} \cdot \hat{\Omega}_{L}). \label{inte}
\end{eqnarray} 
It is expressed as an integral depending on the order $l$ of the Legendre polynomial $P_{l}(\hat{\Omega}_{1} \cdot \hat{\Omega}_{L})$. $K^{\dagger}_{l}(\hat{\Omega})$ can be calculated from the lowest few orders (see \S\ref{sec2} for example). It becomes increasingly difficult to perform the integral as order $l$ increases. Based on the eigenvalues of the density matrix obtained in Fan, \textit{et al.} (2004), Katsura, \textit{et al.} (2007$a$), we make a conjecture on the explicit form of the operator $K^{\dagger}_{l}(\hat{\Omega})$ for generic order $l$:

\textit{Conjecture 1}
	\begin{eqnarray}
	K^{\dagger}_{l}(\hat{\Omega})=\left(\frac{4\pi}{S+1}\right)^{2}I_{l}\left(\frac{1}{2}J(J+1)-\frac{1}{2}S(\frac{1}{2}S+1)\right)A^{\dagger}_{J}. \label{conj2}
\end{eqnarray}
Here the polynomial $I_{l}\left(x\right)$ satisfy the recurrence relation
\begin{eqnarray}
	I_{l+1}(x)=
	\frac{2l+1}{\left(S+l+2\right)^{2}}\left(\frac{4x}{l+1}+l\right)I_{l}\left(x\right)
	-\frac{l}{l+1}\left(\frac{S-l+1}{S+l+2}\right)^{2}I_{l-1}(x) \label{recu}
\end{eqnarray}
with $I_{0}=1$ and $I_{1}=\frac{x}{(\frac{S}{2}+1)^{2}}$.

Note that it is important that $K^{\dagger}_{l}(\hat{\Omega})\propto A^{\dagger}_{J}$ defined in (\ref{aope}) and $I_{l}(x)$ has the same order as the Legendre polynomial $P_{l}(x)$. The recurrence relation (\ref{recu}) was proposed in Freitag \& Muller-Hartmann (1991) and used in Katsura, \textit{et al.} (2007$a$) to obtain the eigenvalues of the density matrix. 
(The original definition of $I_l(x)$ differed from our definition in (\ref{recu}) by a factor of $\frac{2l+1}{4\pi}$.)
\textit{Conjecture 1} (\ref{conj2}) is an alternative form of theorem \ref{theorem1}, %\textbf{Theorem 1},
which also gives eigenvalues through the recurrence relation (\ref{recu}). %We emphasize that \textit{Conjecture 1} together with recurrence relation (\ref{recu}) is a stronger version of \ref{theorem1} %\textbf{Theorem 1}. 
%If we could prove (\ref{conj2}) and (\ref{recu}), then the statement that the set of vectors $\{|\mbox{G}; J, \hat{\Omega}\rangle\}$ (\ref{eige}) are eigenvectors of the density matrix $\boldsymbol{\rho}_{L}$ (\ref{matr}) is completely verified. 
Indeed, expressions (\ref{sum}), altogether with (\ref{conj2}) and (\ref{recu}) yields that
\begin{eqnarray}
	&&\boldsymbol{\rho}_{L}|\mbox{G}; J, \hat{\Omega}\rangle \label{appl} \\ 		&=&\frac{1}{(S+1)^{2}}\sum^{S}_{l=0}(2l+1)\lambda^{L-1}(l,S)
I_{l}\left(\frac{1}{2}J(J+1)-\frac{1}{2}S(\frac{1}{2}S+1)\right)|\mbox{G}; J, \hat{\Omega}\rangle. \nonumber
\end{eqnarray}
Non-zero eigenvalues ($J=0, 1, \ldots, S$) are seen from (\ref{appl}) as
\begin{eqnarray}
	\Lambda(J)\equiv\frac{1}{(S+1)^{2}}\sum^{S}_{l=0}(2l+1)\lambda^{L-1}(l,S)
I_{l}\left(\frac{1}{2}J(J+1)-\frac{1}{2}S(\frac{1}{2}S+1)\right). \label{eiva}
\end{eqnarray}
Since all other eigenvalues of the density matrix are vanishing, then we conclude again that the density matrix $\boldsymbol{\rho}_{L}$ (\ref{matr}) is a projector onto a subspace of dimension $(S+1)^{2}$. This subspace is spanned by the set of vectors $\{|\mbox{G}; J, \hat{\Omega}\rangle\}$ (\ref{eige}). (The rank of the set is equal to $(S+1)^2$.) Furthermore, we observe from (\ref{eiva}) that non-zero eigenvalues $\Lambda(J)$ depend only on $J$, not on $\hat{\Omega}$. Therefore, $\{|\mbox{G}; J, \hat{\Omega}\rangle\}$ with fixed $J$ value spans a degenerate subspace with the same eigenvalue.

%%%%%%%%%%%%%%%%%%%%%%%%%%%%%%%%%%%%%%%%%%%%%%%%%%%%%%%%%%%%%%%%%%%%%%%%%%%%%%%%%%%%%%%%%%%%%%%
\subsection{Eigenvalues of the Density Matrix (Normalization of Degenerate VBS States)}
\label{sec1.6}

Based on the diagonalized form (\ref{fipr}), eigenvalues of the density matrix $\boldsymbol{\rho}_{L}$ can be derived from the normalization of degenerate VBS states. We obtain an explicit expression for eigenvalues in terms of Wigner $3j$-symbols in this subsection. 

First, the following property is important:
Normalization of the degenerate VBS state $|{\rm VBS}_L(J,M)\rangle$ depends only on $J$ and is independent of $M$. With the introduction of total spin operators of the block $S^{\pm}_{\mbox{\scriptsize{tot}}}$, $S^{z}_{\mbox{\scriptsize{tot}}}$ and $\boldsymbol{S^{2}_{\mbox{\scriptsize{tot}}}}$ (see \ref{secA3}), we prove the statement as follows:
\begin{eqnarray}
&& \langle \mbox{VBS}_{L}(J,M\pm 1)|\mbox{VBS}_{L}(J,M\pm 1)\rangle
\nonumber \\
&=& \frac{1}{(J\mp M)(J \pm M+1)}\langle\mbox{VBS}_{L}(J,M)|S^{\mp}_{\mbox{\scriptsize{tot}}}S^{\pm}_{\mbox{\scriptsize{tot}}}|\mbox{VBS}_{L}(J,M)\rangle
\nonumber \\
&=& \frac{1}{(J\mp M)(J \pm M+1)}\langle\mbox{VBS}_{L}(J,M)|(\boldsymbol{S}^{2}_{\mbox{\scriptsize{tot}}}-(S^{z}_{\mbox{\scriptsize{tot}}})^{2}\mp S^{z}_{\mbox{\scriptsize{tot}}})|\mbox{VBS}_{L}(J,M)\rangle
\nonumber \\
&=& \langle\mbox{VBS}_{L}(J,M)|\mbox{VBS}_{L}(J,M)\rangle.
\end{eqnarray}
Here we have used the fact that $|\mbox{VBS}_{L}(J,M)\rangle$ is the eigenstate of $\boldsymbol{S}^{2}_{\mbox{\scriptsize{tot}}}$ and $S^{z}_{\mbox{\scriptsize{tot}}}$ with eigenvalues $J(J+1)$ and $M$, respectively (see \ref{secA3}). 

It is also realized that normalization of $|\mbox{VBS}_{L}(J,M)\rangle$ can be calculated from integrating the inner product of $|\mbox{G};J,{\hat\Omega}\rangle$ with itself over the unit vector $\hat{\Omega}$ such that
\begin{eqnarray}
 &&\frac{1}{4\pi}\int \rd\hat{\Omega}\langle {\rm G};J,\hat{\Omega}| \mbox{G};J,{\hat \Omega}\rangle  \label{novb} \\
%&=& \frac{(S+J+1)!(S-J)!J!J!}{2J+1}\sum_{M=-J}^{J}\sum_{M'=-J}^{J}
%\langle {\rm VBS}_L(J,M)|{\rm VBS}_L(J,M)\rangle 
%\frac{1}{4\pi}\int d{\hat \Omega}X^*_{J,M}X_{J,M'}
%\nonumber \\
&=& \frac{(S+J+1)!(S-J)!J!J!}{(2J+1)!}\langle {\rm VBS}_L(J,M)|{\rm VBS}_L(J,M)\rangle.\nonumber
\end{eqnarray}
In obtaining this relation (\ref{novb}) we have used expansion (\ref{line}) and orthogonality (\ref{orth}) in \ref{secA2}.

Let's consider the integral involved in (\ref{novb}). Using coherent state representation (\ref{cohe}) and completeness relation (\ref{comp}) as before, we obtain
\begin{eqnarray}
&& \frac{1}{4\pi}\int \rd\hat{\Omega}\langle\mbox{G};J,\hat{\Omega}|\mbox{G};J,{\hat\Omega}\rangle
\label{inne} \\
&=& \frac{1}{4\pi}\left[\frac{(2S+1)!}{4\pi}\right]^L \int \rd\hat{\Omega} \int \left[\prod^{L}_{j=1} \rd\hat{\Omega}_{j}\right] \prod^{L-1}_{j=1}\left[\frac{1}{2}(1-\hat{\Omega}_{j}\cdot\hat{\Omega}_{j+1})\right]^{S} \nonumber \\
&\cdot&\left[\frac{1}{2}(1-\hat{\Omega}_{1}\cdot\hat{\Omega}_{L})\right]^{S-J}
\left[\frac{1}{2}(1+\hat{\Omega}_{1} \cdot\hat{\Omega})\right]^{J}
\left[\frac{1}{2}(1+\hat{\Omega} \cdot\hat{\Omega}_{L})\right]^{J}. \nonumber
\end{eqnarray}
Now we expand $\left[\frac{1}{2}(1-\hat{\Omega}_i \cdot\hat{\Omega}_j)\right]^{J}$ in terms of spherical harmonics as in (\ref{expa}),
%Now similar to (\ref{expa}), we expand $\left[\frac{1}{2}(1-\hat{\Omega}_i \cdot\hat{\Omega}_j)\right]^{J}$ in terms of spherical harmonics as
%\begin{eqnarray}
%\left[\frac{1}{2}(1-\hat{\Omega}_i \cdot\hat{\Omega}_j)\right]^{J}
%=\frac{4\pi}{J+1}\sum_{l=0}^J \lambda(l,J)\sum_{m=-l}^l Y_{lm}({\hat\Omega})
%Y^{\ast}_{lm}({\hat\Omega})
%\end{eqnarray}
%with
%\begin{equation}
%\lambda(l,J)=\frac{(-1)^l J!(J+1)!}{(J-l)!(J+l+1)!}.
%\end{equation}
then integrate over ${\hat\Omega}$ and from ${\hat\Omega}_2$ to ${\hat\Omega}_{L-1}$, the right hand side of (\ref{inne}) is equal to
\begin{eqnarray}
&& \frac{4\pi ((2S+1)!)^{L}}{(S+1)^{L-1}(S-J+1)(J+1)^{2}}
\sum_{l_{1}=0}^{S} \sum_{l_{L}=0}^{S-J} \sum_{l=0}^{J}
\sum_{m_{1}=-l_{1}}^{l_{1}} \sum_{m_{L}=-l_{L}}^{l_{L}} \sum_{m=-l}^{l}  \nonumber \\
&\cdot& \int \rd\hat{\Omega}_{1} \int \rd\hat{\Omega}_{L}
\lambda^{L-1}(l_{1},S) \lambda(l_{L},S-J) \lambda^{2}(l,J) \nonumber \\
&\cdot& Y_{l_{1},m_{1}}({\hat\Omega}_{1})Y_{l_{L},m_{L}}({\hat\Omega}_{1})Y_{l,m}({\hat\Omega}_1)
Y^{\ast}_{l_{1},m_{1}}({\hat\Omega}_{L})Y^{\ast}_{l_{L},m_{L}}({\hat\Omega}_{L})Y^{\ast}_{l,m}({\hat\Omega}_{L}).\label{pre1}
\end{eqnarray}
Here we apply the following useful formula:
\begin{eqnarray}
&&\int \rd\hat{\Omega} Y_{l_{1},m_{1}}({\hat\Omega})Y_{l_{L},m_{L}}({\hat\Omega})Y_{l,m}({\hat\Omega})
\nonumber \\
&=&\sqrt{\frac{(2l_1+1)(2l_L+1)(2l+1)}{4\pi}}
\left(\begin{array}{c c c}
l_{1} & l_{L} & l \\
0   &  0  & 0 
\end{array}\right)
\left(\begin{array}{c c c}
l_{1} & l_{L} & l \\
m_{1} & m_{L} & m 
\end{array}\right), \label{wign}
\end{eqnarray}
where $\left(\begin{array}{c c c}
l_{1} & l_{L} & l \\
m_{1} & m_{L} & m 
\end{array}\right)$ is the Wigner $3j$-symbol.
Using formula (\ref{wign}), we carry out the integrals in (\ref{pre1}) and obtain
\begin{eqnarray}
&& \frac{((2S+1)!)^{L}}{(S+1)^{L-1}(S-J+1)(J+1)^{2}} \sum_{l_{1}=0}^S \sum_{l_{L}=0}^{S-J} \sum_{l=0}^J
\sum_{m_{1}=-l_{1}}^{l_{1}} \sum_{m_{L}=-l_{L}}^{l_{L}} \sum_{m=-l}^{l} 
\nonumber \\
&\cdot& (2l_{1}+1)(2l_{L}+1)(2l+1)
\lambda^{L-1}(l_{1},S) \lambda(l_{L},S-J) \lambda^{2}(l,J) \nonumber \\
&\cdot& \left(\begin{array}{c c c}
l_{1} & l_{L} & l \\
0   &  0  & 0 
\end{array}\right)^{2}
\left(\begin{array}{c c c}
l_{1} & l_{L} & l \\
m_{1} & m_{L} & m 
\end{array}\right)^{2}.
\label{wigner}
\end{eqnarray}
The symbols obey the following orthogonality relation:
\begin{eqnarray}
	\sum_{m_{1},m_{L}}(2l+1)
	\left(\begin{array}{c c c}
	l_{1} & l_{L} & l \\
	m_{1} & m_{L} & m 
	\end{array}\right)
	\left(\begin{array}{c c c}
	l_{1} & l_{L} & l^{\prime} \\
	m_{1} & m_{L} & m^{\prime} 
	\end{array}\right)=\delta_{ll^{\prime}} \delta_{mm^{\prime}}. \label{3jor}
\end{eqnarray}
Using this orthogonality (\ref{3jor}), we can recast expression (\ref{wigner}) as
\begin{eqnarray}
&&\frac{((2S+1)!)^{L}}{(S+1)^{L-1}(S-J+1)(J+1)^{2}} \sum_{l_{1}=0}^{S} \sum_{l_{L}=0}^{S-J} \sum_{l=0}^{J} 
\label{pre2} \\
&\cdot&(2l_{1}+1)(2l_{L}+1)(2l+1) \lambda^{L-1}(l_{1},S) \lambda(l_{L},S-J) \lambda^{2}(l,J)
\left(\begin{array}{c c c}
l_{1} & l_{L} & l \\
0   &  0  & 0 
\end{array}\right)^{2}.\nonumber
\end{eqnarray}
The explicit value of $\left(\begin{array}{c c c}
l_{1} & l_{L} & l \\
0   &  0  & 0 
\end{array} \right)$ is given by
\begin{eqnarray}
&&\left(\begin{array}{c c c}
l_{1} & l_{L} & l \\
0   &  0  & 0 
\end{array} \right) \label{3jva} \\
&=&(-1)^{g} \sqrt{\frac{(2g-2l_{1})!(2g-2l_{L})!(2g-2l)!}{(2g+1)!}}\frac{g!}{(g-l_{1})!(g-l_{L})!(g-l)!},
\nonumber
\end{eqnarray}
if $l_{1}+l_{L}+l=2g$ ($g \in {\bf N}$), otherwise zero. 
Finally, normalization of degenerate VBS states $|\mbox{VBS}_{L}(J,M)\rangle$ is obtained as
\begin{eqnarray}
&&\langle\mbox{VBS}_{L}(J,M)|\mbox{VBS}_{L}(J,M)\rangle \label{renovb} \\
&=&\frac{(2J+1)!((2S+1)!)^{L}}{(S+1)^{L-1}(S+J+1)!(S-J+1)!(J+1)!(J+1)!}\sum_{l_{1}=0}^{S} \sum_{l_{L}=0}^{S-J} \sum_{l=0}^{J}
\nonumber \\
&&(2l_{1}+1)(2l_{L}+1)(2l+1) \lambda^{L-1}(l_{1},S) \lambda(l_{L},S-J) \lambda^{2}(l,J)
\left(\begin{array}{c c c}
l_{1} & l_{L} & l \\
0   &  0  & 0 
\end{array}\right)^{2}. \nonumber
\end{eqnarray}

Combining results of (\ref{fipr}) and (\ref{renovb}), we arrive at the following theorem on eigenvalues:
\begin{theorem}
\label{theorem2}
Eigenvalues $\Lambda(J)$ $(J=0,\ldots,S)$ of the density matrix are independent of $\hat{\Omega}$ and/or $M$ in defining eigenvectors (see (\ref{eige}) and (\ref{devb})). An explicit expression is given by the following triple sum
	\begin{eqnarray}
	&&\Lambda(J) \nonumber \\		&=&\left[\frac{S+1}{(2S+1)!}\right]^{L}\frac{S!S!}{S+1}\langle\mbox{VBS}_{L}(J,M)|\mbox{VBS}_{L}(J,M)\rangle \label{eivaex} \\
	&=&\frac{(2J+1)!S!S!}{(S+J+1)!(S-J+1)!(J+1)!(J+1)!}\sum^{S}_{l_{1}=0} \sum^{S-J}_{l_{L}=0} \sum^{J}_{l=0}
	\nonumber \\
	&&\cdot(2l_{1}+1)(2l_{L}+1)(2l+1) \lambda^{L-1}(l_{1},S) \lambda(l_{L},S-J) \lambda^{2}(l,J)
	\left(\begin{array}{ccc}
	l_{1} & l_{L} & l \\
	0   &  0  & 0 
	\end{array}\right)^{2}. \nonumber
	\end{eqnarray}
\end{theorem}

Although not straightforward to verify, this expression (\ref{eivaex}) should be consistent with eigenvalues given through the recurrence expression (\ref{eiva}). We could check the case when $S=1$ that
\begin{eqnarray}
&&\langle\mbox{VBS}_{L}(0,0)|\mbox{VBS}_{L}(0,0)\rangle=\frac{1}{2}(3^{L}+3(-1)^{L}),
\nonumber \\
&&\langle\mbox{VBS}_{L}(1,M)|\mbox{VBS}_{L}(1,M)\rangle=\frac{1}{2}(3^{L}-(-1)^{L}),
\end{eqnarray}
where we have used the selection rule of the Wigner $3j$-symbol. From (\ref{normvbs}) we find that $\langle\mbox{VBS}|\mbox{VBS}\rangle=2\cdot 3^{L}$, so that we obtain the correct eigenvalues of the density matrix from the above result (\ref{eivaex}) (see \S\ref{sec3} for comparison).

We shall emphasize at this point that given eigenvalues (\ref{eivaex}), both von Neumann entropy 
\begin{eqnarray}
	S_{v.N}=-Tr\left[\boldsymbol{\rho_{L}}\ln{\boldsymbol{\rho_{L}}}\right]=-\sum^{S}_{J=0}(2J+1)\Lambda(J)\ln\Lambda(J) \label{vonn}
\end{eqnarray}
and Renyi entropy 
\begin{eqnarray}
	S_{R}=\frac{1}{1-\alpha}\ln \left\{Tr\left[\boldsymbol{\rho}^{\alpha}_{L}\right]\right\}=\frac{1}{1-\alpha}\ln\left\{\sum^{S}_{J=0}(2J+1)\Lambda^{\alpha}(J)\right\} \label{renyi}
\end{eqnarray}
can be expressed directly.

%%%%%%%%%%%%%%%%%%%%%%%%%%%%%%%%%%%%%%%%%%%%%%%%%%%%%%%%%%%%%%%%%%%%%%%%%%%%%%%%%%%%%%%%%%%%%%%%%%%%%%%%%%%%%%%%%%%%%%%%%%%%%%%%%%%%%%%%%%%%%%%%%%%%%%%%%%%%%%%%%%%%%%%%%%%%%%%%%%%%%%%%%%%%%%%%%
\section{Density Matrix in the Large Block Limit}
\label{sec2}

In the limit $L\rightarrow\infty$, that is when the size of the block becomes large, we learned from Fan, \textit{et al.} (2004), Katsura, \textit{et al.} (2007$a$), Vidal, \textit{et al.} (2003), Hadley (2008) that the von Neumann entropy reaches the saturated value $S_{v.N}=\ln\left(S+1\right)^{2}$. Then the density matrix (denoted by $\boldsymbol{\rho}_{\infty}$ in the limit) can only take the form (see Nielsen \& Chuang 2000 for a general proof)
\begin{eqnarray}
	\boldsymbol{\rho}_{\infty}=\frac{1}{(S+1)^{2}}I_{(S+1)^{2}}\oplus \Phi_{\infty}, \label{denl}
\end{eqnarray}
where $I_{(S+1)^{2}}$ is the identity of dimension $(S+1)^{2}$ and $\Phi_{\infty}$ is an infinite dimensional matrix with only zero entries. In this section, we give a proof of \textit{Conjecture 1} (\ref{conj2}) in the limiting case as $L\rightarrow\infty$. Then we shall verify the structure of the density matrix (\ref{denl}) explicitly.

We first realize from (\ref{lamb}) that as $L\rightarrow\infty$, $\lambda^{L-1}(l,S)\rightarrow \delta_{l,0}$. Therefore only the first term with $l=0$ is left in (\ref{sum}). So that we need only to calculate $K^{\dagger}_{0}(\hat{\Omega})$:
\begin{eqnarray}
	K^{\dagger}_{0}(\hat{\Omega})&=&\int \rd\hat{\Omega}_{1} \rd\hat{\Omega}_{L} 
	\left(u_{1}a^{\dagger}_{1}+v_{1}b^{\dagger}_{1}\right)^{S} \left(uu^{\ast}_{1}+vv^{\ast}_{1}\right)^{J}    \nonumber \\ &&\cdot
\left(u^{\ast}_{1}v^{\ast}_{L}-v^{\ast}_{1}u^{\ast}_{L}\right)^{S-J}
\left(uu^{\ast}_{L}+vv^{\ast}_{L}\right)^{J}
\left(u_{L}a^{\dagger}_{L}+v_{L}b^{\dagger}_{L}\right)^{S}. \label{k0}
\end{eqnarray} 

It is useful to know transformation properties of the integrand in (\ref{k0}) under $SU(2)$.
The pair of variables $(u, v)$ defined in (\ref{spin}) and bosonic annihilation operators $(a, b)$ in the Schwinger representation both transform as spinors under $SU(2)$. That is to say, if we take an arbitrary element $D\in SU(2)$ ($2\times 2$ matrix), then $(u, v)$, \textit{etc.} transform according to
\begin{eqnarray}
\left( \begin{array}{c}
	u \\ v 
\end{array}\right)
\rightarrow D
\left( \begin{array}{c}
	u \\ v
\end{array} \right). \label{tran}
\end{eqnarray}
On the other hand, $(u^{\ast}, v^{\ast})$, $(-v, u)$, $(a^{\dagger}, b^{\dagger})$ and $(-b, a)$ transform conjugately to $(u, v)$. That is to say $(u^{\ast}, v^{\ast})$, \textit{etc.} transform according to
\begin{eqnarray}
	\left( \begin{array}{c}
	u^{\ast} \\ v^{\ast} 
\end{array}\right)
\rightarrow D^{\ast}
\left( \begin{array}{c}
	u^{\ast} \\ v^{\ast}
\end{array}\right). \label{ctra}
\end{eqnarray}
The combinations appeared in $K^{\dagger}_{0}(\hat{\Omega})$ (\ref{k0})
\begin{eqnarray}
	u_{1}a^{\dagger}_{1}+v_{1}b^{\dagger}_{1},\ \ uu^{\ast}_{1}+vv^{\ast}_{1},\ \ u^{\ast}_{1}v^{\ast}_{L}-v^{\ast}_{1}u^{\ast}_{L},\ \ uu^{\ast}_{L}+vv^{\ast}_{L},\ \
	u_{L}a^{\dagger}_{L}+v_{L}b^{\dagger}_{L} \label{comb}
\end{eqnarray}
as well as $A^{\dagger}_{J}$ in (\ref{aope}), boundary operator $B^{\dagger}$ in (\ref{bope1}), \textit{etc.} all transform covariantly under $SU(2)$, i.e. those expressions keep their form in the new (transformed) coordinates. 

These transformation properties (\ref{tran}), (\ref{ctra}) can be used to simplify the $K^{\dagger}_{0}(\hat{\Omega})$ integral. We first make a $SU(2)$ transform
\begin{eqnarray}
	D_{u_{L}}=
\left( \begin{array}{cc}
	u^{\ast}_{L} & v^{\ast}_{L} \\
	-v_{L} & u_{L}
\end{array} \right), \qquad
D_{u_{L}}
\left( \begin{array}{c}
	u_{L} \\ v_{L}
\end{array} \right)=\left( \begin{array}{c}
	1 \\ 0
\end{array} \right), \label{utra}
\end{eqnarray}
under the part of the integral (\ref{k0}) over $\hat{\Omega}_{1}$. Then this part of integral  becomes
\begin{eqnarray}
	\int \rd\hat{\Omega}_{1} 
\left(u_{1}a^{\dagger}_{1}+v_{1}b^{\dagger}_{1}\right)^{S} \left(uu^{\ast}_{1}+vv^{\ast}_{1}\right)^{J}  \left(-v^{\ast}_{1}\right)^{S-J}.
\end{eqnarray}
This can be calculated using binomial expansion and the result is
\begin{eqnarray}
	\frac{4\pi}{S+1}\left(ua^{\dagger}_{1}+vb^{\dagger}_{1}\right)^{J}  \left(-b^{\dagger}_{1}\right)^{S-J}. \label{part1}
\end{eqnarray}
Then we make an inverse transform in (\ref{part1}) using $D^{-1}_{u_{L}}=D^{\dagger}_{u_{L}}$, consequently (\ref{k0}) is put in a form with a single integral over $\hat{\Omega}_{L}$ remaining:
\begin{eqnarray}
K^{\dagger}_{0}(\hat{\Omega})&=&\frac{4\pi}{S+1}\left(ua^{\dagger}_{1}+vb^{\dagger}_{1}\right)^{J} \label{rema} \\
&&\cdot\int \rd\hat{\Omega}_{L} 
\left(a^{\dagger}_{1}v^{\ast}_{L}-b^{\dagger}_{1}u^{\ast}_{L}\right)^{S-J}
\left(uu^{\ast}_{L}+vv^{\ast}_{L}\right)^{J}   \left(u_{L}a^{\dagger}_{L}+v_{L}b^{\dagger}_{L}\right)^{S}. \nonumber
\end{eqnarray}
Now we make another $SU(2)$ transform using
\begin{eqnarray}
	D_{u}=
\left( \begin{array}{cc}
	u^{\ast} & v^{\ast} \\
	-v & u
\end{array} \right), \qquad
D_{u}
\left( \begin{array}{c}
	u \\ v
\end{array} \right)=\left( \begin{array}{c}
	1 \\ 0
\end{array} \right),
\end{eqnarray}
then the remaining integral over $\hat{\Omega}_{L}$ in (\ref{rema}) becomes
\begin{eqnarray}
\int \rd\hat{\Omega}_{L} 
\left(a^{\dagger}_{1}v^{\ast}_{L}-b^{\dagger}_{1}u^{\ast}_{L}\right)^{S-J}
\left(u^{\ast}_{L}\right)^{J}   \left(u_{L}a^{\dagger}_{L}+v_{L}b^{\dagger}_{L}\right)^{S}. \label{reml}
\end{eqnarray} 
Using again binomial expansion, this integral (\ref{reml}) yields
\begin{eqnarray}
\frac{4\pi}{S+1}
\left(a^{\dagger}_{1}b^{\dagger}_{L}-b^{\dagger}_{1}a^{\dagger}_{L}\right)^{S-J}
\left(a^{\dagger}_{L}\right)^{J}. \label{part2}
\end{eqnarray} 
At last we make an inverse transform in (\ref{part2}) using $D^{-1}_{u}=D^{\dagger}_{u}$ and plug the result into (\ref{rema}), the final form is
\begin{eqnarray}
K^{\dagger}_{0}(\hat{\Omega})
%=\left(\frac{4\pi}{S+1}\right)^{2}
%\left(ua^{\dagger}_{1}+vb^{\dagger}_{1}\right)^{J}  \left(a^{\dagger}_{1}b^{\dagger}_{L}-b^{\dagger}_{1}a^{\dagger}_{L}\right)^{S-J}
%\left(ua^{\dagger}_{L}+vb^{\dagger}_{L}\right)^{J} \nonumber \\
=\left(\frac{4\pi}{S+1}\right)^{2}A^{\dagger}_{J}. \label{fina}
\end{eqnarray}
This expression is consistent with \textit{Conjecture 1} (\ref{conj2}), which also proves that $\{|\mbox{G}; J, \hat{\Omega}\rangle\}$ is a set of eigenvectors of the density matrix as $L\rightarrow\infty$. Let's denote the density matrix in the limit by $\boldsymbol{\rho}_{\infty}$. Then (\ref{fina}) leads to the result (see (\ref{appl}))
\begin{eqnarray}
	\boldsymbol{\rho}_{\infty}|\mbox{G}; J, \hat{\Omega}\rangle=
	\frac{1}{(S+1)^{2}}|\mbox{G}; J, \hat{\Omega}\rangle. \label{appll}
\end{eqnarray}

We find from (\ref{appll}) that the limiting eigenvalue $\Lambda_{\infty}=\frac{1}{(S+1)^{2}}$ is independent of $J$. Any vector of the $(S+1)^2$-dimensional subspace spanned by the set $\{|\mbox{G}; J, \hat{\Omega}\rangle\}$ is an eigenvector of $\boldsymbol{\rho}_{\infty}$ with the same eigenvalue $\frac{1}{(S+1)^{2}}$. Therefore $\boldsymbol{\rho}_{\infty}$ acts on this subspace as (proportional to) the identity $I_{(S+1)^{2}}$. So that we have proved explicitly that the density matrix takes the form (\ref{denl}) in the large block limit. In addition, we also derive from the eigenvalues that the von Neumann entropy $S_{v.N}=-\sum^{S}_{J=0}(2J+1)\Lambda_{\infty}\ln\Lambda_{\infty}$ coincides with the Renyi entropy $S_{R}=\frac{1}{1-\alpha}\ln\left\{\sum^{S}_{J=0}(2J+1)\Lambda^{\alpha}_{\infty}\right\}$ and is equal to the saturated value $\ln(S+1)^{2}$.

%%%%%%%%%%%%%%%%%%%%%%%%%%%%%%%%%%%%%%%%%%%%%%%%%%%%%%%%%%%%%%%%%%%%%%%%%%%%%%%%%%%%%%%%%%%%%%%%%%%%%%%%%%%%%%%%%%%%%%%%%%%%%%%%%%%%%%%%%%%%%%%%%%%%%%%%%%%%%%%%%%%%%%%%%%%%%%%%%%%%%%%%%%%%%%%%%
\section{Density Matrix for Spin $S=1$}
\label{sec3}

In the case of spin $S=1$, we could prove \textit{Conjectures 1} (\ref{conj2}) for finite block by calculating $K^{\dagger}_{1}(\hat{\Omega})$ defined in (\ref{inte}) using similar methods as been used in \S\ref{sec2}. However, in this special case $S=1$, we have an alternative algebraic proof. We shall use a different representation in which the eigenvectors of the density matrix form an orthogonal basis (maximally entangled states). The formulation is base on Fan, \textit{et al.} (2004).

%%%%%%%%%%%%%%%%%%%%%%%%%%%%%%%%%%%%%%%%%%%%%%%%%%%%%%%%%%%%%%%%%%%%%%%%%%%%%%%%%%%%%%%%%%%%%%%%
\subsection{Ground State of the Unique Hamiltonian}
\label{sec3.1}

The unique Hamiltonian is given by (\ref{uniq1}). In order to represent the unique ground state, we first introduce the following notation for convenience (Fan, \textit{et al.} 2004): 
\begin{eqnarray}
	|\alpha\rangle \equiv (-1)^{1+\delta_{\alpha, 0}}I\otimes\sigma_{\alpha}|0\rangle, \qquad \alpha=0, 1, 2, 3 \label{alph}
\end{eqnarray}
where $\sigma_{0}\equiv I$ (2-dimensional identity), $\sigma_{\alpha=1, 2, 3}$ are Pauli matrices and $|0\rangle\equiv \frac{-1}{\sqrt{2}}(|\uparrow\downarrow\rangle-|\downarrow\uparrow\rangle)$ is the singlet state (antisymmetric projection) of two spin-$1/2$'s. The four states (\ref{alph}) (maximally entangled states) form an orthonormal basis of the Hilbert space of two spin-$1/2$ operators.

The spin-$1$ state at each site is represented by a symmetric projection of two spin-$1/2$ states given by (\ref{alph}) for $\alpha=1, 2, 3$. Let's take the $j$th site for example. The two spin-$1/2$'s are labeled by $(j, \bar{j})$ (from left to right, respectively). Then the spin-$1$ states are prepared by projecting these two spin-$1/2$'s (4-dimensional space) onto a symmetric 3-dimensional subspace spanned by
\begin{eqnarray}
	|1\rangle_{j\bar{j}} &=& \frac{1}{\sqrt{2}}(|\uparrow\rangle_{j}|\uparrow\rangle_{\bar{j}}-|\downarrow\rangle_{j}|\downarrow\rangle_{\bar{j}}), \nonumber \\
	|2\rangle_{j\bar{j}} &=& \frac{-\ri}{\sqrt{2}}(|\uparrow\rangle_{j}|\uparrow\rangle_{\bar{j}}+|\downarrow\rangle_{j}|\downarrow\rangle_{\bar{j}}), \nonumber \\
	|3\rangle_{j\bar{j}} &=& \frac{-1}{\sqrt{2}}(|\uparrow\rangle_{j}|\downarrow\rangle_{\bar{j}}+|\downarrow\rangle_{j}|\uparrow\rangle_{\bar{j}}). \label{3j}
\end{eqnarray}  
Thus the two ending spin-$1/2$'s are labeled as site $\bar{0}$ and $N+1$. The unique ground state in this representation is (Fan, \textit{et al.} 2004; Affleck, \textit{et al.} 1987, 1988)
\begin{eqnarray}
	|\mbox{G}\rangle=\left(\otimes^{N}_{j=1}P_{j\bar{j}}\right)|0\rangle_{\bar{0}1}|0\rangle_{\bar{1}2}\cdots|0\rangle_{\bar{N}N+1}. \label{grou1}
\end{eqnarray}
Here $P_{j\bar{j}}$ projects two spin-$1/2$ states onto a symmetric subspace, which describes spin-$1$. Using basis (\ref{alph}), we have
\begin{eqnarray}
	P_{j\bar{j}}=\sum^{3}_{\alpha=1}|\alpha\rangle_{j\bar{j}}\langle \alpha|. \label{proj}
\end{eqnarray}
A crucial step (see Fan, \textit{et al.} 2004) is that the ground state (\ref{grou1}) can be expressed in a different form using 
\begin{eqnarray}
	|0\rangle_{\bar{A}B}|0\rangle_{\bar{B}C}=\frac{-1}{2}\sum^{3}_{\alpha=0}|\alpha\rangle_{B\bar{B}}\left[I_{\bar{A}}\otimes\left(\sigma_{\alpha}\right)_{C}\right]|0\rangle_{\bar{A}C} \label{maxi}
\end{eqnarray}
for arbitrary labels $A$, $B$ and $C$. Repeatedly using relation (\ref{maxi}), the product of $|0\rangle$'s in (\ref{grou1}) can be rewritten as
\begin{eqnarray}
	&&|0\rangle_{\bar{0}1}|0\rangle_{\bar{1}2}\cdots|0\rangle_{\bar{N}N+1} \label{prod} \\
	&=&\left(\frac{-1}{2}\right)^{N}\sum^{3}_{\alpha_{1}, \cdots, \alpha_{N}=0}|\alpha_{1}\rangle\cdots|\alpha_{N}\rangle\left[I_{\bar{0}}\otimes\left(\sigma_{\alpha_{N}}\cdots\sigma_{\alpha_{1}}\right)_{N+1}\right]|0\rangle_{\bar{0}N+1}. \nonumber
\end{eqnarray}
Then by projecting onto symmetric subspace spanned by $|\alpha=1, 2, 3\rangle$, the ground state (\ref{grou1}) takes the form (Verstraete, \textit{et al.} 2004$a$; Fannes, \textit{et al.} 1992)
\begin{eqnarray}
	|\mbox{G}\rangle=\frac{1}{3^{N/2}}\sum^{3}_{\alpha_{1}, \cdots, \alpha_{N}=1}|\alpha_{1}\rangle\cdots|\alpha_{N}\rangle\left[I_{\bar{0}}\otimes\left(\sigma_{\alpha_{N}}\cdots\sigma_{\alpha_{1}}\right)_{N+1}\right]|0\rangle_{\bar{0}N+1}. \label{grou}
\end{eqnarray}
Note that this ground state (\ref{grou}) is normalized and we have re-written the overall phase for it has no physical content.

%%%%%%%%%%%%%%%%%%%%%%%%%%%%%%%%%%%%%%%%%%%%%%%%%%%%%%%%%%%%%%%%%%%%%%%%%%%%%%%%%%%%%%%%%%%%%%%%
\subsection{Density Matrix of a Block of Bulk Spins}
\label{sec3.2}

Given the ground state in the form (\ref{grou}), we obtain the density matrix of a block of $L$
contiguous spins starting at site $k$ by tracing out spin degrees of freedom outside the block using basis (\ref{alph}):
\begin{eqnarray}
	\boldsymbol{\rho}_{L}\equiv Tr_{\bar{0}, 1, \ldots, k-1, k+L, \ldots, N, N+1} \ |\mbox{G}\rangle\langle \mbox{G}|. \label{den1}
\end{eqnarray}
The result is independent of the starting site $k$ and the total length $N$ (see Fan, \textit{et al.} 2004). We choose $k=1$, $N=L$ so that the density matrix reads (Fan, \textit{et al.} 2004)
\begin{eqnarray}
	\boldsymbol{\rho}_{L} \label{matr1}=
	\frac{1}{3^{L}}\sum^{3}_{\alpha, \alpha^{\prime}=1}|\alpha_{1}\rangle \langle\alpha^{\prime}_{1}|\cdots|\alpha_{L}\rangle \langle\alpha^{\prime}_{L}| \langle 0|I\otimes (\sigma_{\alpha^{\prime}_{1}}\cdots\sigma_{\alpha^{\prime}_{L}})I\otimes (\sigma_{\alpha_{L}}\cdots\sigma_{\alpha_{1}})|0\rangle.%_{\bar{0}L+1}. 
\nonumber\\
\end{eqnarray}

%%%%%%%%%%%%%%%%%%%%%%%%%%%%%%%%%%%%%%%%%%%%%%%%%%%%%%%%%%%%%%%%%%%%%%%%%%%%%%%%%%%%%%%%%%%%%%%%
\subsection{Ground States of the Block Hamiltonian}
\label{sec3.3}

The degenerate Hamiltonian is given by (\ref{dege1}). We choose the length of the spin chain to be equal to that of the block, then the block Hamiltonian $H_{b}\equiv H_{deg}$ with $N=L$ reads
\begin{eqnarray}
	H_{b}=\frac{1}{2}\sum^{L-1}_{j=1} \left(\boldsymbol{S}_{j}\cdot\boldsymbol{S}_{j+1}+\frac{1}{3}\left(\boldsymbol{S}_{j}\cdot\boldsymbol{S}_{j+1}\right)^{2}+\frac{2}{3}\right). \label{degel1}
\end{eqnarray}
Any linear combination of states of the following form
\begin{eqnarray}
	|\mbox{G}; \chi_{1}, \chi_{\bar{L}}\rangle
\equiv\left(\otimes^{L}_{j=1}P_{j\bar{j}}\right)|\chi_{1}\rangle_{1}|0\rangle_{\bar{1}2}|0\rangle_{\bar{2}3}\cdots|0\rangle_{\overline{L-1}L}|\chi_{\bar{L}}\rangle_{\bar{L}} \label{dgini}
\end{eqnarray}
is a ground state of the block Hamiltonian (\ref{degel1}). In (\ref{dgini}) we have made notation $|\chi\rangle \equiv |\uparrow \mbox{or} \downarrow\rangle$ represents the two spin-$1/2$ states and $P_{j\bar{j}}$ is defined in (\ref{proj}). Let's make a particular linear combination of these $|\mbox{G}; \chi_{1}, \chi_{\bar{L}}\rangle$ states using (\ref{alph}) and write the four ($\alpha=0,1,2,3$) linearly independent ground states of the block Hamiltonian (\ref{degel1}) as follows
\begin{eqnarray}
	|\mbox{G}; \alpha\rangle
\equiv\left(\otimes^{L}_{j=1}P_{j\bar{j}}\right)|\alpha\rangle_{\bar{L}1}|0\rangle_{\bar{1}2}|0\rangle_{\bar{2}3}\cdots|0\rangle_{\overline{L-1}L}. \label{dgalph}
\end{eqnarray}
Now we go through the same steps as from (\ref{grou1}) to (\ref{grou}), the resultant form of the four ground states ($\alpha=0, 1, 2, 3$) is
\begin{eqnarray}
	|\mbox{G}; \alpha\rangle=
	\sum^{3}_{\alpha_{1}, \cdots, \alpha_{L}=1}|\alpha_{1}\rangle\cdots|\alpha_{L}\rangle \ \langle\alpha_{L}|\sigma_{\alpha}\otimes\left(\sigma_{\alpha_{L-1}}\cdots\sigma_{\alpha_{1}}\right)|0\rangle.%_{\bar{L}L}. 
	\label{galph}
\end{eqnarray}
Again we have re-written the overall phase for simplicity. These four states are orthogonal, and the normalization is given by
\begin{eqnarray}
	\langle \mbox{G}; \alpha|\mbox{G}; \alpha\rangle=\left\{ \begin{array}{cc}
         \frac{1}{4}(3^{L}+3(-1)^{L}), & \alpha=0;\\ \\
         \frac{1}{4}(3^{L}-(-1)^{L}), & \alpha=1, 2, 3. \end{array}\right. \label{norm}
\end{eqnarray}

%%%%%%%%%%%%%%%%%%%%%%%%%%%%%%%%%%%%%%%%%%%%%%%%%%%%%%%%%%%%%%%%%%%%%%%%%%%%%%%%%%%%%%%%%%%%%%%%
\subsection{Eigenvectors of the Density Matrix}
\label{sec3.4}

According to theorem \ref{theorem1}, %\textbf{Theorem 1},
the degenerate ground states (\ref{galph}) are eigenvectors of the density matrix (\ref{matr1}). Let's apply $\boldsymbol{\rho}_{L}$ to $|\mbox{G}; \alpha\rangle$ and use orthogonality of the $|\alpha\rangle$ states. Then we obtain
\begin{eqnarray}
	\boldsymbol{\rho}_{L}|\mbox{G}; \alpha\rangle=\frac{1}{3^{L}}\sum^{3}_{\alpha_{1}, \cdots, \alpha_{L}=1}|\alpha_{1}\rangle\cdots|\alpha_{L}\rangle \ C_{\alpha_{1}\cdots\alpha_{L}} \label{rhoga}
\end{eqnarray}
with coefficient
\begin{eqnarray}
	C_{\alpha_{1}\cdots\alpha_{L}}&=&\sum^{3}_{\alpha^{\prime}_{1}, \cdots, \alpha^{\prime}_{L}=1}
\langle\alpha^{\prime}_{L}|\sigma_{\alpha}\otimes(\sigma_{\alpha^{\prime}_{L-1}}\cdots\sigma_{\alpha^{\prime}_{1}})|0\rangle \label{coef} \\
	&&%_{\bar{L}L} 
\cdot\langle 0|I\otimes (\sigma_{\alpha^{\prime}_{1}}\cdots\sigma_{\alpha^{\prime}_{L}}) 
I\otimes (\sigma_{\alpha_{L}}\cdots\sigma_{\alpha_{1}})|0\rangle.%_{\bar{0}L+1}. 
\nonumber
\end{eqnarray}
It can be shown by induction that
\begin{eqnarray}
	\sum^{3}_{\alpha^{\prime}_{1}, \cdots, \alpha^{\prime}_{L-1}=1}
(I\otimes\sigma_{\alpha^{\prime}_{L-1}}\cdots\sigma_{\alpha^{\prime}_{1}})|0\rangle
\langle 0|(I\otimes \sigma_{\alpha^{\prime}_{1}}\cdots\sigma_{\alpha^{\prime}_{L-1}})=
\sum^{3}_{\beta=0}A_{\beta}|\beta\rangle\langle \beta| \label{indu}
\end{eqnarray}
with
\begin{eqnarray}
	A_{\beta}=\left\{ \begin{array}{cc}
         \frac{1}{4}(3^{L-1}+3(-1)^{L-1}), & \beta=0;\\ \\
         \frac{1}{4}(3^{L-1}-(-1)^{L-1}), & \beta=1, 2, 3. \end{array}\right. \label{indu2}
\end{eqnarray}
Therefore the coefficient $C_{\alpha_{1}\cdots\alpha_{L}}$ defined in (\ref{coef}) can be simplified as
\begin{eqnarray}
	&&C_{\alpha_{1}\cdots\alpha_{L}} \label{coefind} \\
	&=&\sum^{3}_{\alpha^{\prime}_{L}=1, \beta=0}
	A_{\beta}\langle\alpha^{\prime}_{L}|\sigma_{\alpha}\otimes I|\beta\rangle%_{\bar{L}L}
	\langle \beta|I\otimes (\sigma_{\alpha^{\prime}_{L}}\sigma_{\alpha_{L}}) 
	I\otimes (\sigma_{\alpha_{L-1}}\cdots\sigma_{\alpha_{1}})|0\rangle.%_{\bar{0}L+1}.
\nonumber
\end{eqnarray}
Straightforward calculation using multiplication rules of Pauli matrices shows that (\ref{coefind}) can be further simplified as
\begin{eqnarray}
	&&C_{\alpha_{1}\cdots\alpha_{L}}=
	3A_{1}\delta_{\alpha,0}\langle\alpha_{L}|I\otimes (\sigma_{\alpha_{L-1}}\cdots\sigma_{\alpha_{1}})|0\rangle \label{coefsimp} \\
	&&+(A_{0}+2A_{1})(1-\delta_{\alpha,0})(\delta_{\alpha\alpha_{L}}\langle 0|-\ri\sum^{3}_{\beta=1}\epsilon_{\alpha\alpha_{L}\beta}\langle \beta|)I\otimes(\sigma_{\alpha_{L-1}}\cdots\sigma_{\alpha_{1}})|0\rangle \nonumber
\end{eqnarray}
where $\epsilon_{\alpha\alpha_{L}\beta}$ is the totally antisymmetric tensor of three indices with $\epsilon_{123}=1$. By realizing that 
\begin{eqnarray}
	\delta_{\alpha\alpha_{L}}\langle 0|-\ri\sum^{3}_{\beta=1}\epsilon_{\alpha\alpha_{L}\beta}\langle \beta|=\langle 0|\sigma_{\alpha_{L}}\sigma_{\alpha}\otimes I=\langle \alpha_{L}|\sigma_{\alpha}\otimes I, \label{real}
\end{eqnarray}
we have reached the final form of the coefficient $C_{\alpha_{1}\cdots\alpha_{L}}$ such that
\begin{eqnarray}
	C_{\alpha_{1}\cdots\alpha_{L}}=\left[3A_{1}\delta_{\alpha,0}+(A_{0}+2A_{1})(1-\delta_{\alpha,0})\right]\langle\alpha_{L}|\sigma_{\alpha}\otimes (\sigma_{\alpha_{L-1}}\cdots\sigma_{\alpha_{1}})|0\rangle. \label{codffin}
\end{eqnarray}
As a result, we plug (\ref{codffin}) into (\ref{rhoga}) and find that
\begin{eqnarray}
	\boldsymbol{\rho}_{L}|\mbox{G}; \alpha\rangle&=&\frac{3A_{1}\delta_{\alpha,0}+(A_{0}+2A_{1})(1-\delta_{\alpha,0})}{3^{L}}\label{eigeqn} \\
	&&\cdot\sum^{3}_{\alpha_{1}, \cdots, \alpha_{L}=1}|\alpha_{1}\rangle\cdots|\alpha_{L}\rangle\langle\alpha_{L}|\sigma_{\alpha}\otimes (\sigma_{\alpha_{L-1}}\cdots\sigma_{\alpha_{1}})|0\rangle. \nonumber
\end{eqnarray}
By comparing with (\ref{galph}), we find that (\ref{eigeqn}) is exactly the statement that $|\mbox{G}; \alpha\rangle$ $(\alpha=0, 1, 2, 3)$ are eigenvectors of the density matrix $\boldsymbol{\rho}_{L}$:
\begin{eqnarray}
	\boldsymbol{\rho}_{L}|\mbox{G}; \alpha\rangle=\Lambda_{\alpha}|\mbox{G}; \alpha\rangle, \qquad \alpha=0, 1, 2, 3 \label{eigen}
\end{eqnarray}
with eigenvalues
\begin{eqnarray}
	\Lambda_{\alpha}=\frac{3A_{1}\delta_{\alpha,0}+(A_{0}+2A_{1})(1-\delta_{\alpha,0})}{3^{L}}
	=\left\{ \begin{array}{cc}
         \frac{1}{4}(1+3(-\frac{1}{3})^{L}), & \alpha=0;\\ \\
         \frac{1}{4}(1-(-\frac{1}{3})^{L}), & \alpha=1, 2, 3. \end{array}\right. \label{eigval1}
\end{eqnarray}
These numbers obtained in (\ref{eigval1}) are exactly the eigenvalues found in Fan, \textit{et al.} (2004), Katsura, \textit{et al.} (2007$a$) for spin-$1$, and are consistent with our explicit expression for eigenvalues (\ref{eivaex}).

We can also prove explicitly that any other eigenvectors of $\boldsymbol{\rho}_{L}$ orthogonal to the set $\{|\mbox{G}; \alpha\rangle\}$ have zero eigenvalue. Let's note that a complete basis of the Hilbert space $\boldsymbol{H}_{L}$ of the block of spins can be chosen as
\begin{eqnarray}
	\{|\alpha_{1}\rangle\cdots|\alpha_{L}\rangle\}, \qquad \alpha=1, 2, 3. \label{hlba}
\end{eqnarray}
The subspace $\boldsymbol{H}_{\Lambda}$ with non-zero eigenvalues is panned by $\{|\mbox{G}; \alpha\rangle\}$, as we have already shown. The Hilbert space can be reduced into a direct sum 
\begin{eqnarray}
	\boldsymbol{H}_{L}=\boldsymbol{H}_{\Lambda}\oplus\boldsymbol{H}_{\Phi}. \label{split}
\end{eqnarray}
We will show that the subspace $\boldsymbol{H}_{\Phi}$ orthogonal to $\boldsymbol{H}_{\Lambda}$ is a subspace of vanishing eigenvalues. Mathematically, this means that for an arbitrary basis vector $|\beta_{1}\rangle\cdots|\beta_{L}\rangle$, we shall have
\begin{eqnarray}
\boldsymbol{\rho}_{L}(I_{L}-P_{\Lambda})|\beta_{1}\rangle\cdots|\beta_{L}\rangle=0, \label{zero}
\end{eqnarray}
where $I_{L}$ is the identity of $\boldsymbol{H}_{L}$ and $P_{\Lambda}$ is the projector onto $\boldsymbol{H}_{\Lambda}$:
\begin{eqnarray}
	I_{L}\equiv \sum^{3}_{\alpha_{1},\cdots,\alpha_{L}=1}|\alpha_{1}\rangle\cdots|\alpha_{L}\rangle\langle\alpha_{1}|\cdots\langle\alpha_{L}|, \qquad
	P_{\Lambda}\equiv \sum^{3}_{\alpha=1}\frac{|\mbox{G}; \alpha\rangle\langle\mbox{G}; \alpha|}{\langle \mbox{G}; \alpha|\mbox{G}; \alpha\rangle}. \label{ip}
\end{eqnarray}
By taking expressions (\ref{matr1}), (\ref{ip}), (\ref{eigen}), and realizing that
\begin{eqnarray}
	\sum^{3}_{\alpha=0}\frac{3^{L}\Lambda_{\alpha}}{\langle \mbox{G}; \alpha|\mbox{G}; \alpha\rangle}|\alpha\rangle \langle \alpha|=\sum^{3}_{\alpha=0}|\alpha\rangle \langle \alpha|=I\otimes I \label{iden},
\end{eqnarray}
we find the left hand side of (\ref{zero}) being equal to
\begin{eqnarray}
	&&\boldsymbol{\rho}_{L}(I_{L}-P_{\Lambda})|\beta_{1}\rangle\cdots|\beta_{L}\rangle \label{comm1} \\
	&=&\frac{1}{3^{L}}\sum^{3}_{\alpha_{1}\cdots\alpha_{L}=1}
	|\alpha_{1}\rangle|\cdots|\alpha_{L}\rangle \
\langle 0|[I\otimes(\sigma_{\beta_{1}}\cdots\sigma_{\beta_{L}}), I\otimes(\sigma_{\alpha_{L}}\cdots\sigma_{\alpha_{1}})]|0\rangle. \nonumber
\end{eqnarray}
We use multiplication rules of Pauli matrices to write the two terms within the commutator in (\ref{comm1}) as
\begin{eqnarray}
	I\otimes(\sigma_{\beta_{1}}\cdots\sigma_{\beta_{L}})=\re^{\ri\theta(\beta)}I\otimes\sigma_{\beta}, \qquad \beta=0, 1, 2, 3; \nonumber \\
	I\otimes(\sigma_{\alpha_{L}}\cdots\sigma_{\alpha_{1}})=\re^{\ri\theta(\alpha)}I\otimes\sigma_{\alpha}, \qquad \alpha=0, 1, 2, 3. \label{ab}
\end{eqnarray}
Here $\re^{\ri\theta(\beta)}$ and $\re^{\ri\theta(\alpha)}$ are two phase factors. Then the commutator is
\begin{eqnarray}
	[I\otimes(\sigma_{\beta_{1}}\cdots\sigma_{\beta_{L}}), I\otimes(\sigma_{\alpha_{L}}\cdots\sigma_{\alpha_{1}})]=\re^{\ri(\theta(\beta)+\theta(\alpha))}I\otimes [\sigma_{\beta}, \sigma_{\alpha}]. \label{comm2}
\end{eqnarray}
There are two possibilities: (i) $\alpha=\beta$ or at least one of the two is equal to zero, then $\sigma_{\beta}$ and $\sigma_{\alpha}$ commutes; (ii) $\alpha\neq\beta\neq 0$, then $[\sigma_{\beta}, \sigma_{\alpha}]=2\ri\sum^{3}_{\gamma=1}\epsilon_{\beta\alpha\gamma}\sigma_{\gamma}$, but we still have $\langle 0|I\otimes\sigma_{\gamma}|0\rangle=\langle 0|\gamma\rangle=0$. Therefore, the factor $\langle 0|[I\otimes(\sigma_{\beta_{1}}\cdots\sigma_{\beta_{L}}), I\otimes(\sigma_{\alpha_{L}}\cdots\sigma_{\alpha_{1}})]|0\rangle$ in (\ref{comm1}) is identically zero. So that we have proved (\ref{zero}). Therefore $\boldsymbol{H}_{\Phi}$ is a subspace with only zero eigenvalues.

%%%%%%%%%%%%%%%%%%%%%%%%%%%%%%%%%%%%%%%%%%%%%%%%%%%%%%%%%%%%%%%%%%%%%%%%%%%%%%%%%%%%%%%%%%%%%%%%%%%%%%%%%%%%%%%%%%%%%%%%%%%%%%%%%%%%%%%%%%%%%%%%%%%%%%%%%%%%%%%%%%%%%%%%%%%%%%%%%%%%%%%%%%%%%%%%
\section{A Different Proof of the Theorem on Eigenvectors}
\label{sec4}

It was shown in \S\ref{sec1}$\,\ref{sec1.4}\,$ that the density matrix takes a diagonal form in the basis of zero-energy ground states of the block Hamiltonian (\ref{degel}). In this section, we show the same result by taking a different approach. This alternative proof of theorem \ref{theorem1} %\textbf{Theorem 1}
does not involve coherent state representation. 

Let's start with the ground state of the unique Hamiltonian (\ref{uniq}) with $N=L$:
\begin{eqnarray}
|\mbox{VBS}\rangle&\equiv&\prod_{j=0}^{L}\left(
				      a_j^{\dagger}b_{j+1}^{\dagger}-b_j^{\dagger}a_{j+1}^{\dagger}\right)^{S}|\mbox{vac}\rangle.
\end{eqnarray}
In order to calculate the density matrix $\boldsymbol{\rho}_L=
Tr_{0,L+1}\left[\boldsymbol{\rho}\right]$, where $\boldsymbol{\rho}$ is defined in (\ref{pure}),
we introduce a useful identity:
\begin{eqnarray}
{}_{0,L+1}\langle J,M|\left(|s\rangle_{0,1}\otimes|s\rangle_{L,L+1}\right)
&=&
\frac{(-1)^{S-J+M}}{(S+1)}|J,-M\rangle_{1,L},\label{EigenTrans}
\end{eqnarray}
where $|J,M\rangle_{0,L+1}$ is identical to the spin state defined in (\ref{crea3}) except for
site indices.
$|s\rangle_{i,j}$ in (\ref{EigenTrans}) is the normalized singlet state with $S$ valence bonds defined as \begin{eqnarray}
 |s\rangle_{i,j}&=&\frac{1}{S!\sqrt{S+1}}\left(
a^{\dagger}_{i}b^{\dagger}_{j}-b^{\dagger}_{i}a^{\dagger}_{j}\right)^{S}|\mbox{vac}\rangle_{i}\otimes|\mbox{vac}\rangle_{j}
\nonumber \\ 
&=&\frac{(-1)^{\frac{S}{2}}}{\sqrt{S+1}}\sum_{m=-S/2}^{S/2}(-1)^{m}|S/2,-m \rangle_{i}\otimes|S/2,m \rangle_{j}. \label{singlet}
\end{eqnarray}
Identity (\ref{EigenTrans}) is derived using properties of the singlet state (\ref{singlet}) and Clebsch-Gordan coefficients as follows:
\begin{eqnarray}
&&{}_{0,L+1}\langle J,M|s\rangle_{0,1}|s\rangle_{L,L+1}\nonumber\\
&=&
\sum^{m_{0}+m_{L+1}=M}_{m_{0},m_{L+1}}(J,M|S/2,m_{0};S/2,m_{L+1})
{}_{0}\langle S/2,m_{0}| {}_{L+1}\langle S/2,m_{L+1}|
\nonumber\\&\cdot&
\frac{(-1)^{\frac{S}{2}}}{\sqrt{S+1}}\sum_{m_{1}=-S/2}^{S/2}(-1)^{m_{1}}|S/2,-m_{1}\rangle_{0}|S/2,m_{1}\rangle_{1}
\nonumber\\&\cdot&
\frac{(-1)^{\frac{S}{2}}}{\sqrt{S+1}}\sum_{m_{L}=-S/2}^{S/2}(-1)^{m_{L}}|S/2,-m_{L}\rangle_{L}|S/2,m_{L}\rangle_{L+1}
%\\&=&
%\frac{1}{\sqrt{(S+1)(S+1)}}
%\sum_{m_0,m_{L+1}}\langle J,M|S/2,m_0;S/2,m_{L+1}\rangle
%\nonumber\\&\times&
%\sum_{i=-S/2}^{S/2}
%\sum_{k=-S/2}^{S/2}(-)^{i+k}
%|S/2,-i\rangle_1
%%\\&\otimes&
%|S/2,-k\rangle_L
%\delta_{m_0i}\delta_{m_{L+1}k}
\nonumber\\&=&
\frac{1}{S+1}
\sum^{m_{0}+m_{L+1}=M}_{m_0,m_{L+1}}(-1)^{m_0+m_{L+1}}
(J,M|S/2,m_0;S/2,m_{L+1})
\nonumber\\&\cdot&
|S/2,-m_0\rangle_1
%\\&\otimes&
|S/2,-m_{L+1}\rangle_L.
\end{eqnarray}
Here the Clebsch-Gordan coefficient is defined by
\begin{equation}
(J,M|S/2,m_0;S/2,m_{L+1})= {}_{i,j}\langle J,M|\left(|S/2,m_{0} \rangle_{i}\otimes|S/2,m_{L+1} \rangle_{j}\right).
\end{equation}
Then using the symmetry property of Clebsch-Gordan coefficients
\begin{eqnarray}
(J,M|S/2,m_{0};S/2,m_{L+1})
&=&
%\langle S/2,m_0;S/2,m_{L+1}|J,M\rangle
%\\&=&
(-1)^{S-J}
(J,-M|S/2,-m_{0};S/2,-m_{L+1}), \nonumber \\
\end{eqnarray}
and the completeness of the basis $\{|S/2,m_{0}\rangle_{0}\otimes|S/2,m_{L+1}\rangle_{L+1}\}$,
we obtain the identity (\ref{EigenTrans}).
%\begin{eqnarray}
%&&\langle l',m'|_{0,L+1}|s\rangle_{1,0}\otimes|s\rangle_{L,L+1}\nonumber\\
%&=&
%\frac{1}{S+1}
%\sum_{m_0,m_{L+1}}(-)^{m_0+m_{L+1}}
%(-)^{S-J}
%\nonumber\\&\times&
%|S/2,-m_0\rangle_1
%%\\&\otimes&
%|S/2,-m_{L+1}\rangle_L
%\langle S/2,-m_0;S/2,-m_{L+1}|J,-M\rangle
%\nonumber\\&=&
%\frac{1}{S+1}
%(-)^{S-J+M}
%|J,-M\rangle_{1,L}.
%\end{eqnarray}

With the help of identity (\ref{EigenTrans}), we calculate the partial inner product of the VBS state with the state $|J,M\rangle_{0,L+1}$, which is involved in taking trace of boundary spins. The VBS state $|\mbox{VBS}\rangle$ is decomposed into the bulk part and edge parts, then
making use of (\ref{EigenTrans}), we have
\begin{eqnarray}
&&{}_{0,L+1}\langle J,M|{\rm VBS}\rangle\nonumber\\
&=&
{}_{0,L+1}\langle J,M|
\prod_{j=0}^{L}\left(
				      a_j^{\dagger}b_{j+1}^{\dagger}-b_j^{\dagger}a_{j+1}^{\dagger}\right)^{S}|\mbox{vac}\rangle
\nonumber\\&=& S! (S+1)!
%\prod_{i=1}^{L-1}\left(
%				      a_i^{\dagger}b_{i+1}^{\dagger}-b_i^{\dagger}a_{i+1}^{\dagger}\right)^{S}|0\rangle_{2\cdots L-1}
%\nonumber\\&&\otimes
%\langle J',M'|_{0,L+1}
%\left(
%				      a_0^{\dagger}b_{1}^{\dagger}-b_0^{\dagger}a_{1}^{\dagger}\right)^{S}
%\left(
%				      a_L^{\dagger}b_{L+1}^{\dagger}-b_L^{\dagger}a_{L+1}^{\dagger}\right)^{S}|0\rangle_{0,L+1}
%\nonumber\\
%&=&
\prod_{j=1}^{L-1}\left(
				      a_j^{\dagger}b_{j+1}^{\dagger}-b_j^{\dagger}a_{j+1}^{\dagger}\right)^{S}
%\nonumber\\&&\cdot
{}_{0,L+1}\langle J,M|s\rangle_{0,1} |s\rangle_{L,L+1} |\mbox{vac}\rangle_{2\cdots L-1}
\nonumber\\
&=&(S!)^2
\prod_{j=1}^{L-1}\left(
				      a_j^{\dagger}b_{j+1}^{\dagger}-b_j^{\dagger}a_{j+1}^{\dagger}\right)^{S}
%\nonumber\\&&\cdot
(-1)^{S-J+M}
|J,-M\rangle_{1,L} |\mbox{vac}\rangle_{2\cdots L-1}
\nonumber\\
&=&
(-1)^{S-J+M}(S!)^2
|{\rm VBS}_L(J,-M)\rangle. \label{partial}
\end{eqnarray}
We see that the $(S+1)^2$ degenerate VBS states $|\mbox{VBS}_L(J,M)\rangle$ defined in (\ref{devb}) appear in the partial inner product (\ref{partial}). As discussed in \S\ref{sec1}$\,\ref{sec1.3}\,$, they form a complete set of zero-energy ground states of the block Hamiltonian (\ref{degel}). 
%\begin{eqnarray}
%|{\rm VBS}_L(J,M)\rangle&\equiv&
%\Psi_{JM}^{\dagger}\prod_{i=1}^{L-1}\left(
%					     a_i^{\dagger}b_{i+1}^{\dagger}-b_i^{\dagger}a_{i+1}^{\dagger}\right)^{S}|0\rangle,
%\Psi_{JM}^{\dagger}|{\rm VBS}_L\rangle.
%\label{EdgeState}
%\end{eqnarray}
These states are nothing but the edge states of the subsystem (block).
%Thus, the reduced density matrix of the $L$ contiguous spins is
%explicitly evaluated as 

Now, it is straightforward to evaluate density matrix as
\begin{eqnarray}
Tr_{0,L+1}\left[\boldsymbol{\rho}\right]
&=&
%{\rm Tr}_{0,L+1}\left[
%\frac{|{\rm VBS}_L\rangle\langle{\rm
%		 VBS}_L|}{\langle{\rm VBS}_L|{\rm VBS}_L\rangle}
%\right]
%\nonumber\\&=&
\sum_{J,M}
\frac{{}_{0,L+1}\langle J,M
|\mbox{VBS}\rangle\langle\mbox{VBS}|
 J,M\rangle_{0,L+1}
}{\langle\mbox{VBS}|\mbox{VBS}\rangle}
\nonumber\\&=&
\frac{(S!)^4}{\langle\mbox{VBS}|\mbox{VBS}\rangle}
\sum_{J,M}|\mbox{VBS}_{L}(J,-M)\rangle
\langle\mbox{VBS}_{L}(J,-M)|.
\end{eqnarray}
This expression is identical to (\ref{fipr}) as we change dummy index from $M$ to $-M$. Therefore, in this approach again we arrive at theorem \ref{theorem1} %\textbf{Theorem 1}
that the density matrix is
proportional to a projector onto a subspace spanned by the $(S+1)^2$ ground states of the block Hamiltonian (\ref{degel}). Normalization $\langle{\rm VBS}|{\rm VBS}\rangle$
is given in \ref{secA1}. States $|\mbox{VBS}_{L}(J,M)\rangle$ are shown to be mutually orthogonal in \ref{secA3}. 
%From this fact, we have now the stronger result than {\it Conjecture 1} (\ref{conj2}), i.e., the set of states $\{ |{\rm VBS}_L(J,M)\rangle ,J=0, 1, ..., S, M= -J, ..., J\}$ are exact eigenvectors of the density matrix $\boldsymbol{\rho}_L$ (\ref{matr}).

%%%%%%%%%%%%%%%%%%%%%%%%%%%%%%%%%%%%%%%%%%%%%%%%%%%%%%%%%%%%%%%%%%%%%%%%%%%%%%%%%%%%%%%%%%%%%%%%%%%%%%%%%%%%%%%%%%%%%%%%%%%%%%%%%%%%%%%%%%%%%%%%%%%%%%%%%%%%%%%%%%%%%%%%%%%%%%%%%%%%%%%%%%%%%%%%
\section{Conclusion}
\label{sec5}

We have studied the density matrix $\boldsymbol{\rho}_{L}$ of a block of $L$ contiguous bulk spins in the AKLT model. The unique Hamiltonian for generic spin-$S$ is given by (\ref{uniq}), which has a unique ground state described by the VBS state (\ref{vbs}) in the Schwinger representation. The density matrix $\boldsymbol{\rho}_{L}$ (\ref{matr}) of the block is obtained by taking trace (\ref{trac}) of all spin degrees of freedom outside the block. The structure of the density matrix has been investigated both for finite and infinite blocks.

For generic spin-$S$ and finite block, two mathematically rigorous results have been established as theorem \ref{theorem1} and \ref{theorem2}. %\textbf{Theorem 1} and \textbf{2}.
In theorem \ref{theorem1} %\textbf{Theorem 1}
we constructed eigenvectors of the density matrix with non-zero eigenvalues. These eigenvectors $|\mbox{G}; J, \hat{\Omega}\rangle$ defined in (\ref{eige}), or $|\mbox{VBS}_{L}(J,M)\rangle$ defined in (\ref{devb}) equivalently, are proved to be the $(S+1)^{2}$ zero-energy ground states of the block Hamiltonian (\ref{degel}). The corresponding eigenvalues are obtained in two different forms. Using nonorthogonal basis $\{|\mbox{G}; J, \hat{\Omega}\rangle\}$, the eigenvalues are given through \textsl{Conjecture 1} (\ref{conj2}) and the recurrence relation (\ref{recu}); while using orthogonal basis $\{|\mbox{VBS}_{L}(J,M)\rangle\}$, in theorem  \ref{theorem2} %\textbf{Theorem 2}
an explicit expression (\ref{eivaex}) for eigenvalues in terms of Wigner $3j$-symbols is derived. Non-zero eigenvalues $\Lambda(J)$ with $J=0, 1, \ldots, S$ ((\ref{eiva}) and (\ref{eivaex})) depend only on $J$ and are independent of $\hat{\Omega}$ and/or $M$ in defining eigenvectors. The density matrix (\ref{fipr}) is a projector onto the subspace of dimension $(S+1)^{2}$ spanned by the set of eigenvectors $\{|\mbox{G}; J, \hat{\Omega}\rangle\}$ and/or $\{|\mbox{VBS}_{L}(J,M)\rangle\}$.

In the large block limit $L\rightarrow\infty$, \textit{Conjecture 1} (\ref{conj2}) is proved and all non-zero eigenvalues $\Lambda_{\infty}$ become the same (\ref{appll}). The infinite dimensional density matrix $\boldsymbol{\rho}_{\infty}$ (\ref{denl}) is a projector onto a $(S+1)^{2}$-dimensional subspace in which it is proportional to the identity. The von Neumann entropy $S_{v.N}$ coincides with the Renyi entropy $S_{R}$ and is equal to the saturated value $\ln(S+1)^{2}$. In the limit the Renyi entropy is $\alpha$ independent, which behaves quite differently from the XY model where the Renyi entropy has an essential singularity as a function of $\alpha$ (see Its, \textit{et al.} 2005; Franchini, \textit{et al.} 2007, 2008).
  
We have also investigated the structure of the density matrix in a special case when spin $S=1$. Both theorem \ref{theorem1} and \ref{theorem2} %\textbf{Theorem 1} and \textbf{2}
are proved using a different representation (\ref{galph}) (maximally entangled states) where all four eigenvectors $|\mbox{G}, \alpha\rangle$ are orthogonal. We have also shown (\ref{zero}) explicitly that any vector orthogonal to the subspace spanned by the set $\{|\mbox{G}, \alpha\rangle\}$ has zero eigenvalue.

Based on the main results obtained in this paper, we end our conclusion by making the following conjecture:

\textit{Conjecture 2}

The structure of the density matrix as a projector onto a subspace is generalizable to inhomogeneous AKLT spin chains (spin values at different lattice sites could be different) and lattices of higher dimensions\footnote{It may even be generalizable to certain classes of arbitrary graphs if the limiting sub-graph can be defined properly.}. In the large block limit, the density matrix should behave as the identity operator within the subspace. i.e. Eq. (\ref{enlim}) is valid for arbitrary large lattices.

%%%%%%%%%%%%%%%%%%%%%%%%%%%%%%%%%%%%%%%%%%%%%%%%%%%%%%%%%%%%%%%%%%%%%%%%%%%%%%%%%%%%%%%%%%%%%%%%%%%%%%%%%%%%%%%%%%%%%%%%%%%%%%%%%%%%%%%%%%%%%%%%%%%%%%%%%%%%%%%%%%%%%%%%%%%%%%%%%%%%%%%%%%%%%%%%
\begin{acknowledgements}
The authors would like to thank Professor Heng Fan, Professor Anatol N. Kirillov and Professor Sergey Bravyi for valuable discussions and suggestions. The work is supported by NSF Grant DMS-0503712 and the Japan Society for the Promotion of Science.
\end{acknowledgements}

%%%%%%%%%%%%%%%%%%%%%%%%%%%%%%%%%%%%%%%%%%%%%%%%%%%%%%%%%%%%%%%%%%%%%%%%%%%%%%%%%%%%%%%%%%%%%%%%%%%%%%%%%%%%%%%%%%%%%%%%%%%%%%%%%%%%%%%%%%%%%%%%%%%%%%%%%%%%%%%%%%%%%%%%%%%%%%%%%%%%%%%%%%%%%%%%
\appendix{Normalization of the VBS State}
\label{secA1}

The VBS state $|\mbox{VBS}\rangle$ (also known to be the ground state of the unique Hamiltonian (\ref{uniq})) defined in (\ref{vbs}) is not normalized. Using the coherent state formalism (\ref{cohe}) and the completeness relation (\ref{comp}), we express the norm square as
\begin{eqnarray}
	&&\langle \mbox{VBS}|\mbox{VBS}\rangle \label{norm1} \\
	&=&\left[\frac{(S+1)!}{4\pi}\right]^{2}\left[\frac{(2S+1)!}{4\pi}\right]^{N}\int \left(\prod^{N+1}_{j=0}\rd\hat{\Omega}_{j}\right)\prod^{N}_{j=0}\left[\frac{1}{2}(1-\hat{\Omega}_{j}\cdot\hat{\Omega}_{j+1})\right]^{S} \nonumber \\	
\end{eqnarray}
where we have used
\begin{eqnarray}
	\langle 0|a^{S+l}b^{S-l}|\hat{\Omega}\rangle=\sqrt{(2S)!}u^{S+l}v^{S-l}.
\end{eqnarray}
Now we expand $\left[\frac{1}{2}(1-\hat{\Omega}_{j}\cdot\hat{\Omega}_{j+1})\right]^{S}$ in terms of spherical harmonics as in (\ref{expa}), then integrate from $\hat{\Omega}_{0}$ to $\hat{\Omega}_{N+1}$. We notice by using the orthogonality of spherical harmonics that each integral contributes a factor of $\frac{4\pi}{S+1}$ except the last one. For example,
\begin{eqnarray}
	&&\int \rd\hat{\Omega}_{0}\left[\frac{1}{2}(1-\hat{\Omega}_{0}\cdot\hat{\Omega}_{1})\right]^{S} \nonumber \\			&=&\frac{4\pi}{S+1}\sum^{S}_{l=0}\lambda(l,S)\sum^{l}_{m=-l}\sqrt{4\pi}Y^{\ast}_{lm}(\hat{\Omega}_{1})\int \rd\hat{\Omega}_{0}Y_{lm}(\hat{\Omega}_{0})Y^{\ast}_{00}(\hat{\Omega}_{0}) 
\nonumber \\
	&=&\frac{4\pi}{S+1}\sqrt{4\pi}Y^{\ast}_{00}(\hat{\Omega}_{1})= \frac{4\pi}{S+1}. \label{exam}
\end{eqnarray}
The last integral over $\hat{\Omega}_{N+1}$ contributes simply a factor of $4\pi$. Consequently, the norm square (\ref{norm1}) is equal to
\begin{eqnarray}
	\langle \mbox{VBS}|\mbox{VBS}\rangle=\left[\frac{(2S+1)!}{S+1}\right]^{N}S!(S+1)!. \label{normvbs}
\end{eqnarray}

%%%%%%%%%%%%%%%%%%%%%%%%%%%%%%%%%%%%%%%%%%%%%%%%%%%%%%%%%%%%%%%%%%%%%%%%%%%%%%%%%%%%%%%%%%%%%%%%
%%%%%%%%%%%%%%%%%%%%%%%%%%%%%%%%%%%%%%%%%%%%%%%%%%%%%%%%%%%%%%%%%%%%%%%%%%%%%%%%%%%%%%%%%%%%%%%%
\appendix{Rank of the Set $\{|\mbox{G}; J, \hat{\Omega}\rangle\}$ with Fixed $J$ Value}
\label{secA2}

For notational convenience, we define
\begin{eqnarray}
	X_{JM}\equiv\frac{u^{J+M}v^{J-M}}{\sqrt{(J+M)!(J-M)!}}, \qquad
	\psi^{\dagger}_{Sm}\equiv\frac{(a^{\dagger})^{S+m}(b^{\dagger})^{S-m}}{\sqrt{(S+m)!(S-m)!}}. \label{2def}
\end{eqnarray}
These two variables transform conjugately with respect to one another under $SU(2)$. $X_{JM}$ has the following orthogonality relation
\begin{eqnarray}
	\int \rd\hat{\Omega}X^{\ast}_{JM}X_{JM^{\prime}}=\frac{4\pi}{(2J+1)!}\delta_{MM^{\prime}}. \label{orth}
\end{eqnarray}
$\psi^{\dagger}_{Sm}$ is a spin state creation operator such that
\begin{eqnarray}
	\psi^{\dagger}_{Sm}|\mbox{vac}\rangle=|S, m\rangle. \label{crea1}
\end{eqnarray}
The operator $A^{\dagger}_{J}$ defined in (\ref{aope}) can be expanded as (see Hamermesh 1989)
\begin{eqnarray}
	A^{\dagger}_{J}&=&\sqrt{\frac{(S+J+1)!(S-J)!J!J!}{2J+1}} \label{aexp} \\
	&&\cdot\sum^{J}_{M=-J}X_{JM}\sum^{m_{1}+m_{L}=M}_{m_{1}, m_{L}}(S/2, m_{1}; S/2, m_{2}|J, M)\ \psi^{\dagger}_{S/2,m_{1}}\otimes\psi^{\dagger}_{S/2,m_{L}}, \nonumber
	\label{expandAJ}
\end{eqnarray}
%(The above eq. seems to be incorrect. 
%\begin{eqnarray}
%	A^{\dagger}_{J}&=&\sqrt{\frac{(S+J+1)!(S-J)!J!J!}{2J+1}} \label{aexp} \\
%	&&\cdot\sum^{J}_{M=-J}X_{JM}\sum^{m_{1}+m_{L}=M}_{m_{1}, m_{L}}(S/2, m_{1}; S/2, m_{2}|J, M)\ \psi^{\dagger}_{S/2,m_{1}}\otimes\psi^{\dagger}_{S/2,m_{L}}, \nonumber
%	\label{expandAJ}
%\end{eqnarray}
%)
where $(S/2, m_{1}; S/2, m_{2}|J, M)$ are the Clebsch-Gordan coefficients. Note that \\ $\psi^{\dagger}_{S/2,m_{1}}$ and $\psi^{\dagger}_{S/2,m_{L}}$ are defined in the Hilbert spaces of spins at site $1$ and site $L$, respectively. We realize that the particular form of the sum over $m_{1}$ and $m_{L}$ in (\ref{aexp}) can be identified as a single spin state creation operator
\begin{eqnarray}
	\Psi^{\dagger}_{JM}\equiv\sum^{m_{1}+m_{L}=M}_{m_{1}, m_{L}}(S/2, m_{1}; S/2, m_{2}|J, M)\ \psi^{\dagger}_{S/2,m_{1}}\otimes\psi^{\dagger}_{S/2,m_{L}}. \label{crea2}
\end{eqnarray}
This operator $\Psi^{\dagger}_{JM}$ acts on the direct product of two Hilbert spaces of spins at site $1$ and site $L$. It has the property that
\begin{eqnarray}
	\Psi^{\dagger}_{JM}|\mbox{vac}\rangle_{1}\otimes|\mbox{vac}\rangle_{L}=|J, M\rangle_{1,L}. \label{crea3}
\end{eqnarray}
Now we can derive the completeness relation of the set $\{|\mbox{G}; J, \hat{\Omega}\rangle\}$ using (\ref{orth}), (\ref{aexp}) and (\ref{crea2}):
\begin{eqnarray}
	&&\int \rd\hat{\Omega}|\mbox{G}; J, \hat{\Omega}\rangle\langle \mbox{G}; J, \hat{\Omega}| \label{compga} \\
	&=&\frac{4\pi}{(2J+1)!}\frac{(S+J+1)!(S-J)!J!J!}{2J+1}\sum^{J}_{M=-J}\Psi^{\dagger}_{JM}|\mbox{VBS}_{L}\rangle\langle \mbox{VBS}_{L}|\Psi_{JM}. \nonumber
\end{eqnarray}
The set of states $\{\Psi^{\dagger}_{JM}|\mbox{VBS}_{L}\rangle, \quad M=-J, \ldots, J\}$  are linearly independent. So that the rank of $\{|\mbox{G}; J, \hat{\Omega}\rangle\}$ with fixed $J$ value is $2J+1$. With the introduction of degenerate VBS states $|\mbox{VBS}_{L}(J, M)\rangle$ in (\ref{devb}), $|\mbox{G};J,\hat{\Omega}\rangle$ can be written as a linear superposition:
\begin{eqnarray}
	|\mbox{G};J,\hat{\Omega}\rangle=\sqrt{\frac{(S+J+1)!(S-J)!J!J!}{2J+1}}\sum^{J}_{M=-J}X_{JM}|\mbox{VBS}_{L}(J, M)\rangle. \label{line}
\end{eqnarray}
More details can be found in Hamermesh (1989).

%%%%%%%%%%%%%%%%%%%%%%%%%%%%%%%%%%%%%%%%%%%%%%%%%%%%%%%%%%%%%%%%%%%%%%%%%%%%%%%%%%%%%%%%%%%%%%%%
%%%%%%%%%%%%%%%%%%%%%%%%%%%%%%%%%%%%%%%%%%%%%%%%%%%%%%%%%%%%%%%%%%%%%%%%%%%%%%%%%%%%%%%%%%%%%%%%
\appendix{Orthogonality of Degenerate VBS States} %$|\mbox{VBS}_L(J,M)\rangle$}
\label{secA3}

The set of degenerate VBS states $\{|\mbox{VBS}_L(J,M)\rangle, J=0,...,S, M=-J,...,J\}$ introduced in (\ref{devb}) are mutually orthogonal. To show this, 
it is convenient to introduce the total spin operators of the subsystem:
\begin{equation}
S^{+}_{\mbox{\scriptsize{tot}}}=\sum^{L}_{j=1} a^{\dagger}_{j} b_{j}, \qquad
S^{-}_{\mbox{\scriptsize{tot}}}=\sum^{L}_{j=1} b^{\dagger}_{j} a_{j}, \qquad
S^{z}_{\mbox{\scriptsize{tot}}}=\sum^{L}_{j=1} (a^{\dagger}_{j} a_{j}-b^{\dagger}_{j} b_{j})/2.
\end{equation}
First we show that the set of operators $\{ S^{+}_{\mbox{\scriptsize{tot}}}, S^{-}_{\mbox{\scriptsize{tot}}}, S^{z}_{\mbox{\scriptsize{tot}}} \}$ commute with the product of valence bonds, i.e.
\begin{equation}
[S^{\pm}_{\mbox{\scriptsize{tot}}}, \prod_{j=1}^{L-1} (a^{\dagger}_{j} b^{\dagger}_{j+1}-b^{\dagger}_{j} a^{\dagger}_{j+1})^{S}]=0, \quad 
[S^{z}_{\mbox{\scriptsize{tot}}}, \prod_{j=1}^{L-1} (a^{\dagger}_{j} b^{\dagger}_{j+1}-b^{\dagger}_{j} a^{\dagger}_{j+1})^{S} ]=0.
\label{commute}
\end{equation}
These commutation relations (\ref{commute}) can be shown in similar ways. Take the commutator with $S^{+}_{\mbox{\scriptsize{tot}}}$ first. We re-write the commutator as
\begin{eqnarray}
&&[S^{+}_{\mbox{\scriptsize{tot}}}, \prod_{j=1}^{L-1} (a^{\dagger}_{j} b^{\dagger}_{j+1}-b^{\dagger}_{j} a^{\dagger}_{j+1})^{S}] \nonumber \\
&=& \sum_{j=1}^{L-1} (a^{\dagger}_{1} b^{\dagger}_{2}-b^{\dagger}_{1} a^{\dagger}_{2})^{S} \cdots 
[S^{+}_{\mbox{\scriptsize{tot}}}, (a^{\dagger}_{j} b^{\dagger}_{j+1}-b^{\dagger}_{j} a^{\dagger}_{j+1})^{S}] \cdots (a^{\dagger}_{L-1} b^{\dagger}_{L}-b^{\dagger}_{L-1} a^{\dagger}_{L})^{S} \nonumber \\
&=& \sum_{j=1}^{L-1} (a^{\dagger}_{1} b^{\dagger}_{2}-b^{\dagger}_{1} a^{\dagger}_{2})^{S} \cdots 
[S^{+}_{j} + S^{+}_{j+1}, (a^{\dagger}_{j} b^{\dagger}_{j+1}-b^{\dagger}_{j} a^{\dagger}_{j+1})^{S}]\cdots \nonumber \\ 
&&\cdots (a^{\dagger}_{L-1} b^{\dagger}_{L}-b^{\dagger}_{L-1} a^{\dagger}_{L})^{S}. \nonumber \\
\end{eqnarray}
Then using commutators $[a_{i}, a^{\dagger}_{j}]=\delta_{ij}$ and $[b_{i}, b^{\dagger}_{j}]=\delta_{ij}$, we find that
\begin{eqnarray}
&& [S^{+}_{j} + S^{+}_{j+1}, (a^{\dagger}_{j} b^{\dagger}_{j+1}-b^{\dagger}_{j} a^{\dagger}_{j+1})^{S}]
\nonumber \\
&=& [a^{\dagger}_{j} b_{j}+a^{\dagger}_{j+1} b_{j+1}, (a^{\dagger}_{j} b^{\dagger}_{j+1}-b^{\dagger}_{j} a^{\dagger}_{j+1})^{S}] \nonumber \\
&=& a^{\dagger}_{j} [b_{j}, (a^{\dagger}_{j} b^{\dagger}_{j+1}-b^{\dagger}_{j} a^{\dagger}_{j+1})^{S}]+
      a^{\dagger}_{j+1} [b_{j+1}, (a^{\dagger}_{j} b^{\dagger}_{j+1}-b^{\dagger}_{j} a^{\dagger}_{j+1})^{S}]
      \nonumber \\
&=& a^{\dagger}_{j} (-S)a^{\dagger}_{j+1} (a^{\dagger}_{j} b^{\dagger}_{j+1}-b^{\dagger}_{j} a^{\dagger}_{j+1})^{S-1}
    +a^{\dagger}_{j+1} S a^{\dagger}_{j} (a^{\dagger}_{j} b^{\dagger}_{j+1}-b^{\dagger}_{j} a^{\dagger}_{j+1})^{S-1}
\nonumber \\
&=&0. \label{comm}
\end{eqnarray}
Therefore $[S^{+}_{\mbox{\scriptsize{tot}}}, \prod_{j=1}^{L-1} (a^{\dagger}_{j} b^{\dagger}_{j+1}-b^{\dagger}_{j} a^{\dagger}_{j+1})^{S}]=0$. In (\ref{comm}) we have used $[b_{j}, (a^{\dagger}_{j} b^{\dagger}_{j+1}-b^{\dagger}_{j} a^{\dagger}_{j+1})^{S}]=-S a^{\dagger}_{j+1} (a^{\dagger}_{j} b^{\dagger}_{j+1}-b^{\dagger}_{j} a^{\dagger}_{j+1})^{S-1}$. 
In a parallel way, we find that the commutator with $S^{-}_{\mbox{\scriptsize{tot}}}$ also vanishes. Next we consider the commutator with $S^{z}_{\mbox{\scriptsize{tot}}}$:
\begin{eqnarray}
&&[S^{z}_{\mbox{\scriptsize{tot}}}, \prod_{j=1}^{L-1} (a^{\dagger}_{j} b^{\dagger}_{j+1}-b^{\dagger}_{j} a^{\dagger}_{j+1})^{S}] \nonumber \\
&=& \sum_{j=1}^{L-1} (a^{\dagger}_{1} b^{\dagger}_{2}-b^{\dagger}_{1} a^{\dagger}_{2})^{S}
\cdots [S^{z}_{j} + S^{z}_{j+1}, (a^{\dagger}_{j} b^{\dagger}_{j+1}-b^{\dagger}_{j} a^{\dagger}_{j+1})^{S}]\cdots \nonumber \\
&&\cdots (a^{\dagger}_{L-1} b^{\dagger}_{L}-b^{\dagger}_{L-1} a^{\dagger}_{L})^{S}. \nonumber \\
\label{comSz}
\end{eqnarray}
In the right hand side of (\ref{comSz}), the commutator involved also vanishes because
\begin{eqnarray}
&&[S^{z}_{j} + S^{z}_{j+1}, (a^{\dagger}_{j} b^{\dagger}_{j+1}-b^{\dagger}_{j} a^{\dagger}_{j+1})^{S}] 
\nonumber \\
&=& \frac{1}{2}[a^{\dagger}_{j} a_{j} -b^{\dagger}_{j} b_{j} +a^{\dagger}_{j+1} a_{j+1} -b^{\dagger}_{j+1} b_{j+1}, (a^{\dagger}_{j} b^{\dagger}_{j+1}-b^{\dagger}_{j} a^{\dagger}_{j+1})^{S}] 
\nonumber \\
&=& a^{\dagger}_{j}  [a_{j}, (a^{\dagger}_{j} b^{\dagger}_{j+1}-b^{\dagger}_{j} a^{\dagger}_{j+1})^{S}]
     -b^{\dagger}_{j}  [b_{j} ,(a^{\dagger}_{j} b^{\dagger}_{j+1}-b^{\dagger}_{j} a^{\dagger}_{j+1})^{S}] 
\nonumber \\
&+& a^{\dagger}_{j+1} [a_{j+1}, (a^{\dagger}_{j} b^{\dagger}_{j+1}-b^{\dagger}_{j} a^{\dagger}_{j+1})^{S}]
     -b^{\dagger}_{j+1} [b_{j+1}, (a^{\dagger}_{j} b^{\dagger}_{j+1}-b^{\dagger}_{j} a^{\dagger}_{j+1})^{S}]
\nonumber \\
&=& 0  \label{com1}
\end{eqnarray}
Substituting (\ref{com1}) into (\ref{comSz}), we obtain $[S^{z}_{\mbox{\scriptsize{tot}}}, \prod_{j=1}^{L-1} (a^{\dagger}_{j} b^{\dagger}_{j+1}-b^{\dagger}_{j} a^{\dagger}_{j+1})^{S}]=0$.
Now we shall show that the state $|\mbox{VBS}_L(J,M)\rangle$ is an eigenstate of $S^{z}_{\mbox{\scriptsize{tot}}}$ and the square of the total spin $\boldsymbol{S}^2_{\mbox{\scriptsize{tot}}}=\frac{1}{2}(S^{+}_{\mbox{\scriptsize{tot}}}S^{-}_{\mbox{\scriptsize{tot}}}+S^{-}_{\mbox{\scriptsize{tot}}}S^{+}_{\mbox{\scriptsize{tot}}}) + (S^{z}_{\mbox{\scriptsize{tot}}})^{2}$ with eigenvalues $M$ and $J(J+1)$, respectively. 
Using the commutation relations (\ref{commute}), we can show that
\begin{eqnarray}
S^{\pm}_{\mbox{\scriptsize{tot}}}|\mbox{VBS}_L(J,M)\rangle = \prod_{j=1}^{L-1} (a^{\dagger}_{j} b^{\dagger}_{j+1}-b^{\dagger}_{j} a^{\dagger}_{j+1})^{S} (S^{\pm}_{1} + S^{\pm}_{L})|J,M \rangle_{1,L} |\mbox{vac}\rangle_{2, ..., L-1} \nonumber \\
S^{z}_{\mbox{\scriptsize{tot}}}|\mbox{VBS}_L(J,M)\rangle = \prod_{j=1}^{L-1} (a^{\dagger}_{j} b^{\dagger}_{j+1}-b^{\dagger}_{j} a^{\dagger}_{j+1})^{S} (S^{z}_{1} + S^{z}_{L})|J,M \rangle_{1,L} |\mbox{vac}\rangle_{2,..., L-1}. \nonumber \\
\end{eqnarray}
Then from the definition of the state $|\mbox{VBS}_L(J,M)\rangle$ and the following relations:
\begin{eqnarray}
(S^{+}_1+S^{+}_{L})|J,M \rangle_{1,L}&=& \sqrt{(J \mp M) (J \pm M +1)} |J, M\pm1 \rangle, 
\nonumber \\
(S^{z}_{1}+S^{z}_{L})|J,M \rangle_{1,L}\rangle &=& M |J,M \rangle_{1,L},
\end{eqnarray}
we obtain 
\begin{eqnarray}
S^{\pm}_{\mbox{\scriptsize{tot}}}|\mbox{VBS}_L(J,M)\rangle &=& \sqrt{(J \mp M) (J \pm M +1)} |\mbox{VBS}_L(J,M\pm1)\rangle, \nonumber \\
S^z_{\mbox{\scriptsize{tot}}}|\mbox{VBS}_L(J,M)\rangle &=& M |\mbox{VBS}_L(J,M)\rangle
\end{eqnarray}
and hence $\boldsymbol{S}^{2}_{\mbox{\scriptsize{tot}}}|\mbox{VBS}_L(J,M)\rangle=J(J+1)|\mbox{VBS}_L(J,M)\rangle$.
It is now proved that $|\mbox{VBS}_L(J,M)\rangle$ is an eigenstate of $S^{z}_{\mbox{\scriptsize{tot}}}$ and $\boldsymbol{S}^{2}_{\mbox{\scriptsize{tot}}}$ with eigenvalues $M$ and $J(J+1)$, respectively. Therefore the states with different eigenvalues $(J, M)$ are orthogonal to each other.

%To show this, it is convenient to introduce the following generating function:
%\begin{equation}
%{\cal G}(J,{\hat \Omega}|J',{\hat \Omega}') \equiv \langle G;J,{\hat \Omega }| G; J',{\hat \Omega}' \rangle.
%\end{equation}
%Using the above generating function, the inner product of the states $|{\rm VBS}_L(J,M)\rangle$ and $|{\rm VBS}_L(J',M)\rangle$ is given by
%\begin{equation}
%\langle {\rm VBS}_L(J,M)|{\rm VBS}_L(J',M)\rangle
%\propto
%\Big(\frac{\partial}{\partial v^*}\Big)^{J-M}
%\Big(\frac{\partial}{\partial v'}\Big)^{J'-M}
%{\cal G}(J,{\hat \Omega}|J',{\hat \Omega}') \bigg|_{u'=u^*=1 \atop v'=v^*=0}
%\end{equation}
%This can be derived from (\ref{expandAJ}). Let us now consider the generating function itself. The integral form of ${\cal G}(J,{\hat \Omega}|J',{\hat \Omega}')$ is the followig form:
%\begin{eqnarray}
%&&{\cal G}(J,{\hat \Omega}|J',{\hat \Omega}')
%\nonumber\\
%&=&
%\Big[\frac{(2S+1)!}{4\pi} \Big]^L
%\int \Big(\prod_{j=1}^L d{\hat \Omega}_j \Big)
%(u^* u_1+v^* v_1)^J (u_1 v_L-v_1 u_L)^{S-J} (u^* u_L+v^* v_L)^J
%\nonumber \\
%&\cdot& (u' u^*_1+v' v^*_1)^{J'} (u^*_1 v^*_L-v^*_1 u^*_L)^{S-J'} (u' u-*_L+v v^*_L)^{J'} \prod_{j=1}^{L-1}|u_j v_{j+1}-v_j u_{j+1}|^S.
%\nonumber \\
%\label{intgenerating}
%\end{eqnarray}
%Using the orthogonality relation of $X_{JM}$ (\ref{orth}) combining with the binomial expansion, 

%%%%%%%%%%%%%%%%%%%%%%%%%%%%%%%%%%%%%%%%%%%%%%%%%%%%%%%%%%%%%%%%%%%%%%%%%%%%%%%%%%%%%%%%%%%%%%%%
%%%%%%%%%%%%%%%%%%%%%%%%%%%%%%%%%%%%%%%%%%%%%%%%%%%%%%%%%%%%%%%%%%%%%%%%%%%%%%%%%%%%%%%%%%%%%%%%
\appendix{Density Matrix and Correlation Functions}
\label{secA4}
The relation between the density matrix and correlation functions was sudied in Jin \& Korepin (2004), Katsura, \textit{et al} (2007$a$), Arovas, \textit{et al} (1988).
It was shown in Section 2 of Jin \& Korepin (2004) that the density matrix contains information of all correlation functions in the ground state. The original proof was for spin $S=1/2$. In this appendix we shall generalize this result to generic spin-$S$ which is applicable to our AKLT model. 

The Hilbert space associated with a spin-$S$ is $(2S+1)$-dimensional. Therefore we could choose a basis of $(2S+1)^{2}$ linearly independent matrices such that an arbitrary operator defined in the Hilbert space can be written as a superposition over the basis. Let's denote the basis by $\{A_{ab}; \ a,b=1,\ldots,2S+1\}$, in which each matrix $A_{ab}$ is labeled by a pair of indices $a$ and $b$ with totally $(2S+1)^{2}$ possible combinations. The matrix element is defined as
\begin{eqnarray}
	(A_{ab})_{kl}=\delta_{ak}\delta_{bl}, \qquad k,l=1,\ldots,2S+1. \label{aab}
\end{eqnarray}
In addition to $\{A_{ab}\}$, we introduce an equivalent ``conjugate'' basis $\{\bar{A}_{ab}\}$ such that
\begin{eqnarray}
	(\bar{A}_{ab})_{kl}=\delta_{al}\delta_{bk}, \qquad a,b,k,l=1,\ldots,2S+1. \label{abar}
\end{eqnarray}
These matrices (\ref{aab}) and (\ref{abar}) are actually matrix representation of operators $\{|S,m\rangle\langle S,m^{\prime}|;\ m,m^{\prime}=-S,\ldots,S\}$. They are normalized such that
\begin{eqnarray}
	Tr(\bar{A}_{ab}A_{cd})=\sum_{k,l}(\bar{A}_{ab})_{kl}(A_{cd})_{lk}=\sum_{k,l}\delta_{al}\delta_{bk}\delta_{cl}\delta_{dk}=\delta_{ac}\delta_{bd}. \label{traceaa}
\end{eqnarray}
Here $Tr$ takes trace at one and the same site. Because of the completeness of $\{A_{ab}\}$ at each site, the density matrix of the block can be written as (see (\ref{den1}))
\begin{eqnarray}
	\boldsymbol{\rho}_{\mbox{\scriptsize{block}}}=Tr_{\mbox{\scriptsize{outside}}}|\mbox{G}\rangle\langle \mbox{G}|=\sum_{\{a_{j}b_{j}\}}\left(\otimes_{j\in\{\mbox{\scriptsize{block}}\}}A_{a_{j}b_{j}}\right)\mbox{coeff}\{a_{j}b_{j}\}, \label{rhoblock}
\end{eqnarray}
where $|\mbox{G}\rangle$ denotes the unique ground state, $Tr_{\mbox{\scriptsize{outside}}}$ takes traces of sites outside the block and $\mbox{coeff}\{a_{j}b_{j}\}$ denotes the coefficient. Using the normalization property (\ref{traceaa}), the coefficient $\mbox{coeff}\{a_{j}b_{j}\}$ with label $j$ taking values within the block can be expressed as
\begin{eqnarray}
	\mbox{coeff}\{a_{j}b_{j}\}&=&\sum_{\{c_{j}d_{j}\}}\prod_{j\in\mbox{\scriptsize{block}}}Tr(\bar{A}_{a_{j}b_{j}}A_{c_{j}d_{j}})\mbox{coeff}\{c_{j}d_{j}\} \nonumber \\
	&=&Tr_{\mbox{\scriptsize{block}}}\left[\left(\otimes_{j\in\mbox{\scriptsize{block}}}\bar{A}_{a_{j}b_{j}}\right)\boldsymbol{\rho}_{\mbox{\scriptsize{block}}}\right] \nonumber \\
&=&Tr_{\mbox{\scriptsize{all}}}\left[\left(\otimes_{j\in\mbox{\scriptsize{block}}}\bar{A}_{a_{j}b_{j}}\right)|\mbox{G}\rangle\langle \mbox{G}|\right]\nonumber \\
&=&\langle \mbox{G}|\left(\otimes_{j\in\mbox{\scriptsize{block}}}\bar{A}_{a_{j}b_{j}}\right)|\mbox{G}\rangle. \label{coeajbj}
\end{eqnarray}
Here $Tr_{\mbox{\scriptsize{block}}}$ takes traces of sites within the block and $Tr_{\mbox{\scriptsize{all}}}$ takes traces of all lattice sites. Combing (\ref{rhoblock}) with (\ref{coeajbj}), we have the final form
\begin{eqnarray}
	\boldsymbol{\rho}_{\mbox{\scriptsize{block}}}=\sum_{\{a_{j}b_{j}\}}\left(\otimes_{j\in\{\mbox{\scriptsize{block}}\}}A_{a_{j}b_{j}}\right)\langle \mbox{G}|\left(\otimes_{j\in\mbox{\scriptsize{block}}}\bar{A}_{a_{j}b_{j}}\right)|\mbox{G}\rangle. \label{rhocorr}
\end{eqnarray}
This is the expression of the density matrix with entries related to multi-point correlation functions $\langle \mbox{G}|\left(\otimes_{j\in\mbox{\scriptsize{block}}}\bar{A}_{a_{j}b_{j}}\right)|\mbox{G}\rangle$ in the ground state. All possible combinations $\{a_{j}b_{j}\}$ are involved in the summation. Therefore, we have prove for generic spin-$S$ that the density matrix contains information of all correlation functions.

%%%%%%%%%%%%%%%%%%%%%%%%%%%%%%%%%%%%%%%%%%%%%%%%%%%%%%%%%%%%%%%%%%%%%%%%%%%%%%%%%%%%%%%%%%%%%%%%%%%%%%%%%%%%%%%%%%%%%%%%%%%%%%%%%%%%%%%%%%%%%%%%%%%%%%%%%%%%%%%%%%%%%%%%%%%%%%%%%%%%%%%%%%%%%%%%

\label{lastpage}
\end{document}